\documentclass[%
 reprint,
showpacs,preprintnumbers,
showkeys,
 amsmath,amssymb,
 aps,
]{revtex4-1}

\usepackage[english]{babel}
\usepackage[utf8]{inputenc}
\usepackage{graphicx}
\usepackage{bm}
\usepackage{hyperref}
\usepackage{color}
\usepackage{epsfig}
\usepackage{pst-grad}
\usepackage{pst-plot}
\usepackage{amsmath}
\usepackage{amsfonts}
\usepackage{amssymb}
\usepackage{amsthm}

\usepackage{include/braket} 
\usepackage{include/dsfont} 

\hypersetup{
    unicode=false,                 
    pdftoolbar=true,               
    pdfmenubar=true,               
    pdffitwindow=false,            
    pdfstartview={FitH},           
    pdftitle={},                   
    pdfauthor={Nicolai Lang},      
    pdfsubject={Quantum physics},  
    pdfcreator={Nicolai Lang},     
    pdfproducer={Nicolai Lang},    
    pdfkeywords={},                
    pdfnewwindow=true,             
    colorlinks=true,               
    linkcolor=blue,                
    citecolor=blue,                
    filecolor=magenta,             
    urlcolor=blue                  
}

\renewcommand{\vec}[1]{\mathbf{#1}}

\newtheorem{Theorem}{Theorem}
\newtheorem{Definition}{Definition}
\newtheorem{Lemma}{Lemma}
\newtheorem{Proposition}{Proposition}

\newcommand{\infig}[4]{\protect\raisebox{#1 mm}{\protect\hspace{#2 mm}\protect\includegraphics[scale=#3]{#4}}}
\newcommand{\comment}[1]{}                  
\newcommand{\bigslant}[2]{{\raisebox{.2em}{$#1$}\left/\raisebox{-.2em}{$#2$}\right.}}

\graphicspath{{./media/}}

\begin{document}


\title{Minimal instances for toric code ground states}

\author{Nicolai Lang}
\email{nicolai@itp3.uni-stuttgart.de}
\author{Hans Peter Büchler}
\affiliation{Institute for Theoretical Physics III, 
  Universität Stuttgart, 70550 Stuttgart, Germany}

\date{\today}

\begin{abstract}
  A decade ago Kitaev's toric code model established the new paradigm of topological quantum 
  computation. Due to remarkable theoretical and experimental progress, the quantum simulation 
  of such complex many-body systems is now within the realms of possibility. 
  Here we consider the question, to which extent the ground states of \textit{small} toric 
  code systems differ from LU-equivalent graph states. We argue that simplistic 
  (though experimentally attractive) setups obliterate the differences between the toric code 
  and equivalent graph states; hence we search for the smallest setups on the square- 
  and triangular lattice, such that the quasi-locality of the toric code hamiltonian becomes 
  a distinctive feature. To this end, a purely geometric procedure to transform a given 
  toric code setup into an LC-equivalent graph state is derived. In combination with an 
  algorithmic computation of LC-equivalent graph states, we find the smallest non-trivial 
  setup on the square lattice to contain $5$ plaquettes and $16$ qubits; on the triangular
  lattice the number of plaquettes and qubits is reduced to $4$ and $9$, respectively.
\end{abstract}



\maketitle



The quest for topological quantum computation \cite{1_nayak_NAATQC} relies on well
controllable quantum systems exhibiting
anyonic statistics.  Several theoretical studies have proposed
microscopic models featuring such
topological phases; the most prominent example being the toric code \cite{1_kitaev_FTQCBA_A}.
While it is generally accepted that the fractional quantum Hall effect is well
described as a topological state of matter 
\cite{1_laughling_AQHE,1_halperin_SQHFQHS,1_arovas_FSQHE_A}, alternative setups such
as cold atomic gases \cite{2_weimer_RQS_A} and ion traps \cite{3_barreiro_OSQSWTI_A}
are considered as suitable building blocks for a quantum simulation of
topological matter.
However, such systems naturally appear with a small number of involved
particles. Regarding actual implementations it is therefore a fundamental
question, how many particles the simulated system must involve in
order to observe the characteristic properties of the system in question. 
In this letter, we address this question for the toric code and present setups
with a minimal number of qubits in order to obtain non-trivial ground states.


In the last years great experimental progress led to the generation
of highly entangled multi-qubit states in realistic settings, such as
cold atomic gases in optical lattices~\cite{2_bloch_QCEUAOL,2_bloch_MBPUG} 
, Rydberg states~\cite{3_urban_ORBBTA,3_gaetan_OCETIARBR,3_saffman_RSMQGEPNA}
, trapped ions~\cite{3_barreiro_OSQSWTI_A,3_home_CMSSITQI,2_blatt_ESTAI,2_haeffner_QCTI} 
and polarized photons~\cite{3_lu_EESPGS,3_ma_QSWPFHSS,3_yao_OEPE}.
On the theoretical side, a powerful tool in characterizing and
describing states with mutilpartite entanglement is given by the
concept of graph states \cite{1_hein_MPEIGS_A,1_hein_EIGSAIA_S},
and several realistic setups for the generation of graph states have been proposed~
\cite{2_clark_EGGSQC,2_kay_GSPQCGAOL,2_bodiya_SGGSETGLO,2_lin_STQPGEPGSE}.

In contrast to the coherent and measurement based state preparation
above, the concept of quantum simulation~\cite{1_feynman_SPWC,1_lloyd_UQS} of complex 
many-body Hamiltonians provides an alternative approach for
the generation of topological phases as a suitable resource for topological quantum 
computation; the paradigm of which has been established by Kitaev's toric 
code \cite{1_kitaev_FTQCBA_A}. 
Subsequently several setups for the quantum simulation of the latter were
discussed~\cite{2_weimer_RQS_A,2_herdman_SGTP,2_mueller_SOQSMBISP,2_weimer_DQSRA}. 
Special focus was laid on the observation of mutual anyonic statistics for the elementary 
excitations \cite{2_han_SDFSAESM,2_paredes_MITMOP,2_jiang_AIPMASL,2_aguado_CMDANAAOL},
and experimental success within the realm of photonic quantum simulation was reported \cite{3_lu_DAFSSQQS,3_pachos_RAFTCQS}.

\begin{figure}[tbp]
  \small
  A\includegraphics[scale=0.56]{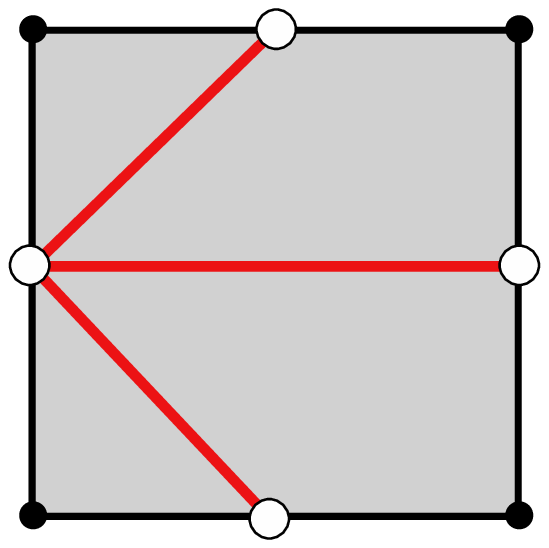}
  \hspace{4pt}
  B\includegraphics[scale=0.30]{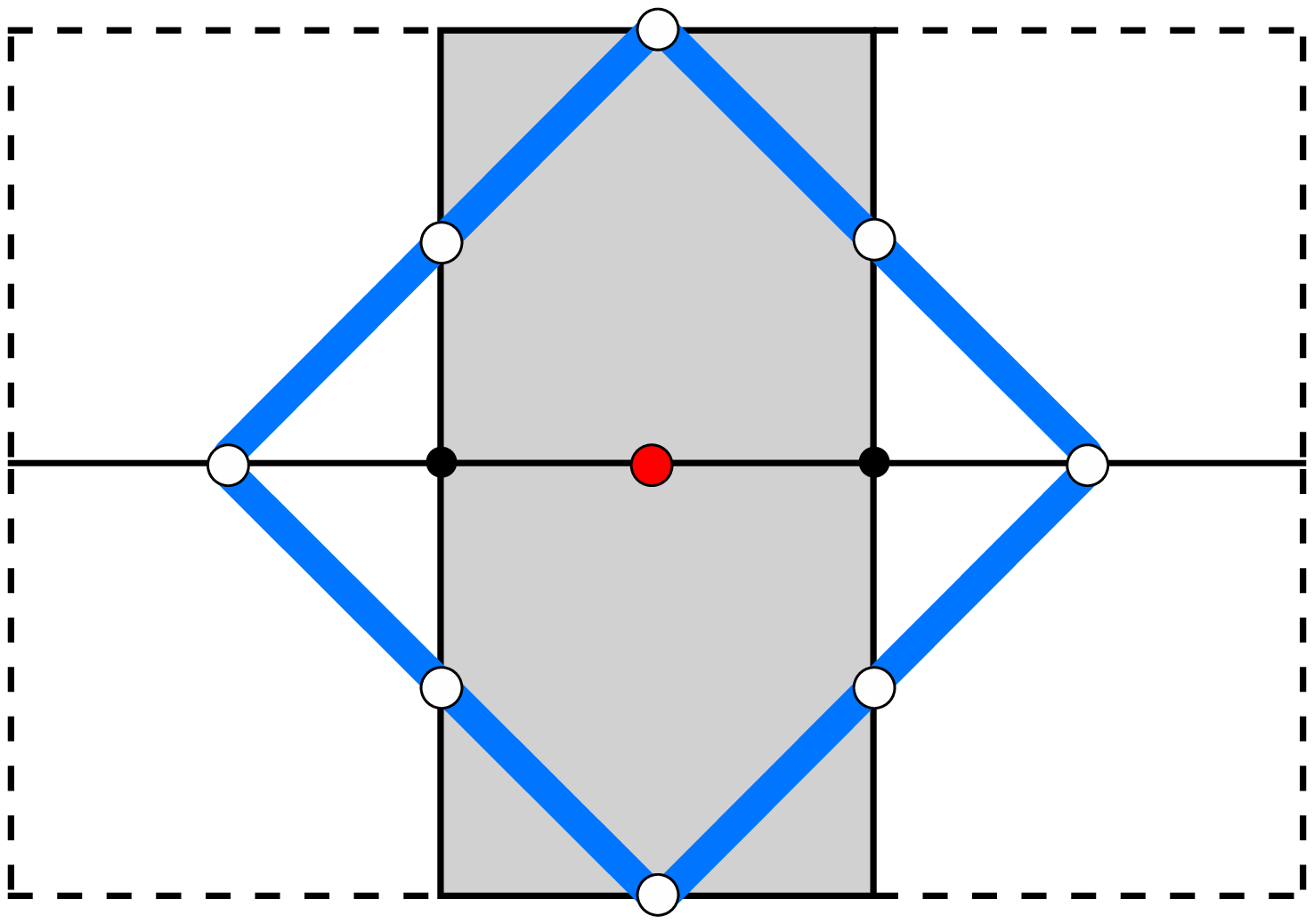}
  \normalsize
  \caption{\label{fig:intro}(Color online) 
    \textbf{A:}~The (non-degenerate) ground state of a single toric code plaquette 
    is a GHZ state and equivalent to the shown graph state.
    Qubits (\infig{-0.7}{-0.2}{0.33}{qubit}) are located on the edges. 
    Vertices (\infig{-0.7}{-0.2}{0.45}{vertex}) denote star operators $A_s$ and
    grey faces plaquette operators $B_p$ acting on the adjacent qubits. 
    The LC-equivalent graph state connects the qubits directly (\infig{0.6}{0.0}{0.50}{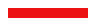}).
    \textbf{B:}~The $8$ neighbours (\infig{-0.5}{0}{0.33}{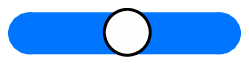}) of the 
    centered qubit (\infig{-0.7}{-0.2}{0.35}{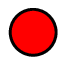}) in the toric code on a square lattice.
    Neighbours and the centered qubit have a star and/or plaquette operator in common.
  }
\end{figure}

However, the systems considered so far are small and their ground states tend to be well 
known multi-qubit states, e.g., the ground state of a single toric code plaquette is a 
simple GHZ state~\cite{1_greenberger_GBBT}, as shown in Fig.~\ref{fig:intro}~(A).
Hence we argue that a proper characterization of toric code systems implies the comparison 
of its ground states with other known multi-qubit states, such as graph states.

\begin{figure*}[tbp]
  A\includegraphics[width=0.25\linewidth]{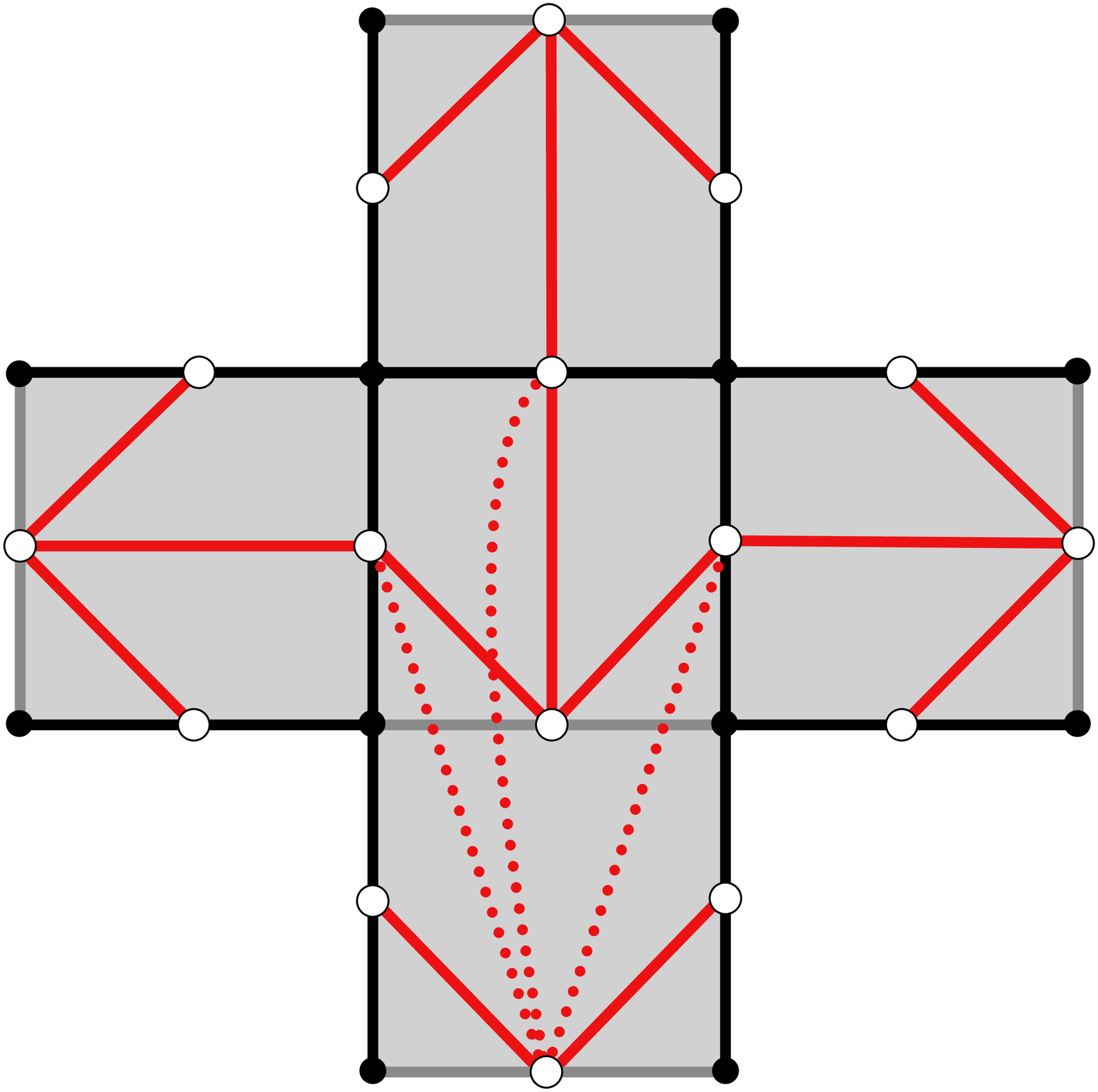}\hspace{20pt}
  B\includegraphics[width=0.25\linewidth]{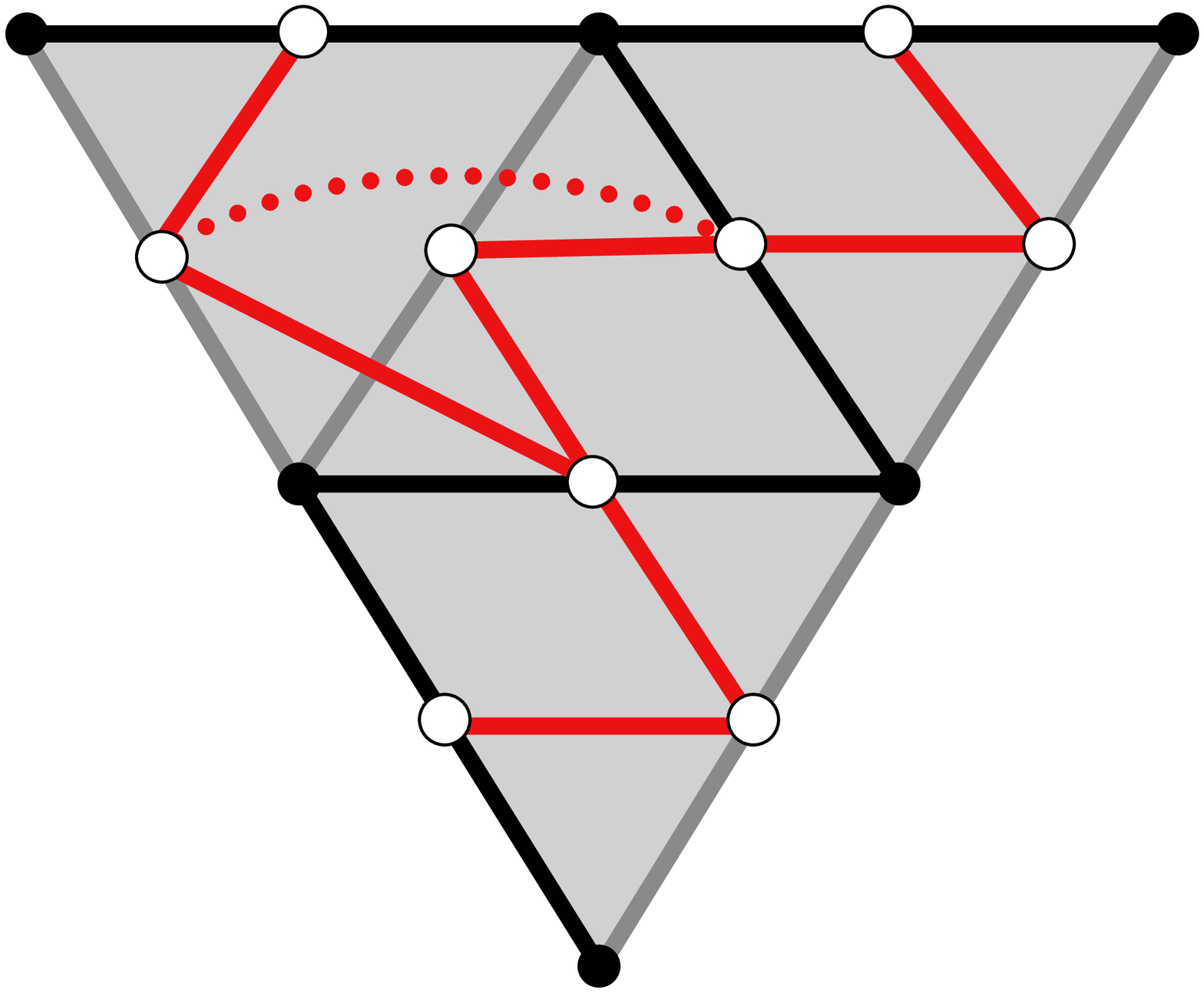}\hspace{20pt}
  C\includegraphics[width=0.25\linewidth]{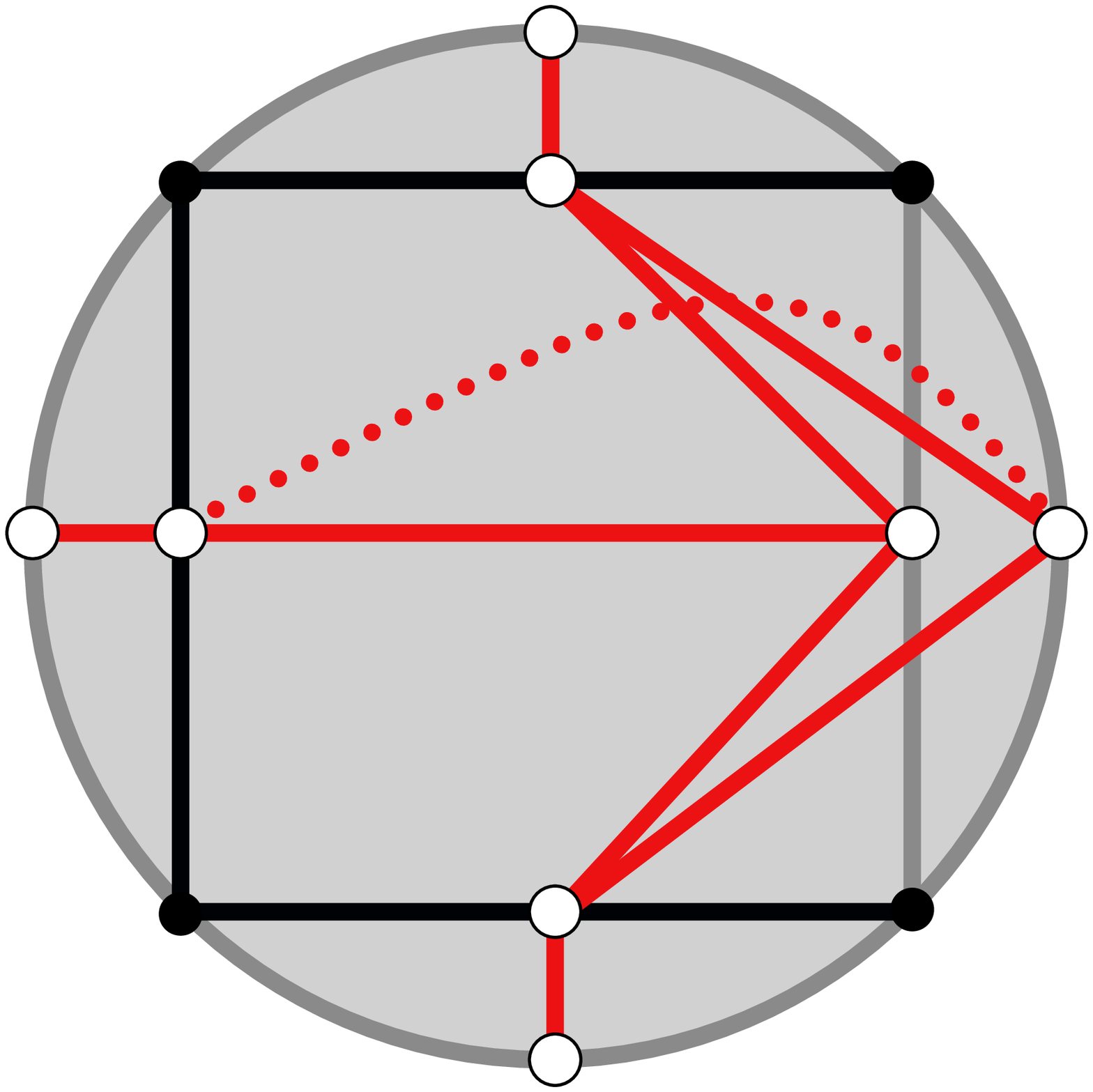}
  \caption{\label{fig:mintcm}(Color online) 
    The minimal \textit{non-local} toric code setup on a square lattice (\textbf{A}) and
    on a triangular lattice (\textbf{B}) with LC-equivalent graph states.
    The system depicted in (\textbf{C}) was used to prove the non-locality of the square 
    lattice setup in (\textbf{A}). \textit{Notation:}~Qubits (\infig{-0.7}{-0.2}{0.33}{qubit}) 
    are located on the edges. Vertices (\infig{-0.7}{-0.2}{0.45}{vertex}) denote star 
    operators $A_s$ and grey faces plaquette operators $B_p$ acting on the adjacent qubits. 
    The drawn LC-equivalent graph states connect qubits directly via \textit{local} 
    (~\infig{0.6}{0.0}{0.50}{red_line}~) and \textit{non-local} 
    (\infig{-0.2}{-0.0}{0.30}{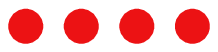}) edges. The graph states are based on the
    black spanning trees (~\infig{0.6}{0.0}{0.50}{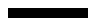}~), see 
    Definition~\ref{def:map} and Theorem~\ref{thm:1} for details.}
\end{figure*}

In this letter, we address the question on the minimal system required
to observe such characteristic properties of the toric code.
Let us explain this in more detail:
The implementation of the toric code naturally gives rise to the
concept of \textit{quasilocal} \footnote{To wit, multi-qubit operations with support 
  that does not scale with the system size.} interactions and nearest neighbour sites:
The qubits are located on the links of a graph which is embedded into a surface, see 
Fig.~\ref{fig:intro}~(B). The qubits connected to the same vertex $s$ give rise to the 
quasilocal stabilizer $A_{s}= \prod_{i \in s} \sigma_{i}^{x}$,
while the qubits surrounding a plaquette $p$ are involved in the many-body 
stabilizer $B_{p} = \prod_{i \in p} \sigma_{i}^{z}$;
the latter are qubits connected to a common vertex of the dual graph. This leads to a
natural definition of nearest-neighbour (or \textit{vicinal}) qubits in the toric code,
as qubits involved in the fundamental stabilizers $A_{s}$ or $B_{p}$.
For a square lattice, the number of nearest neighbours is therefore
eight, see Fig.~\ref{fig:intro}~(B). It is a fundamental property of the toric
code, that it is generated by \textit{quasilocal} stabilizer operators, and
therefore, its experimental realization requires only \textit{quasilocal} interactions.

As mentioned above, for small systems the ground state of the toric code is equivalent to
well known states: for example, the non-degenerate ground state of the setup with
four qubits depicted in Fig.~\ref{fig:intro}~(A) is equivalent to a GHZ state. 
Here, and throughout the manuscript, two states are termed ``equivalent'',
if they differ only by a sequence of single qubit rotations, that is, local unitary operations.
On the other hand, GHZ states belong to the class of 
graph states \cite{1_hein_MPEIGS_A,1_hein_EIGSAIA_S}
which can be generated efficiently by applying two qubit operations 
(see Refs.~\cite{1_hein_MPEIGS_A,2_cabello_OPGS} and Chapter ~\ref{sec:preliminaries}).
As a stabilizer quantum code the toric code ground states are \textit{always} equivalent 
to a class of graph states.
For the above example [Fig.~\ref{fig:intro}~(A)], the depicted equivalent graph state exhibits 
only connections (i.e. edges) between qubits
which are \textit{vicinal} within the toric code. However, this condition will break 
down for large setups: graph states, which are
equivalent to the toric code ground states, will in general exhibit connections 
between qubits that are \textit{not} nearest-neighbours, and therefore their 
experimental realization requires non-local
interactions. Hence a fundamental property of the toric code is the 
\textit{quasilocal} nature of its stabilizers
-- which distinguishes it from the equivalent \textit{non-local} graph states. 
Many applications of the toric code as topologically protected quantum memory rely 
on this local character.

A minimal setup for topological matter should inherit this crucial property.
In this manuscript, we present minimal setups for the toric code, such that the equivalent
graph states are \textit{non-local}, to wit, they require connections between qubits that are
\textit{not} nearest-neighbours. A major obstacle is that the equivalence
between the toric code ground states and graph states is not unique, since 
(in general) there are multiple equivalent graph states.
As a consequence, we have to exclude the existence of \textit{any} equivalent graph state
with only local connections. We therefore first derive a simple, geometric procedure to construct
equivalent graph states from a given toric code setup.
In a second step, all equivalent graph states for promising candidates are generated and
tested on their property of non-local connections.

The results for the most important setups are shown in Fig.~\ref{fig:mintcm}:
We find that for the common square lattice the minimal number of qubits is $N=16$, 
see Fig.~\ref{fig:mintcm}~(A), while
for a triangular setup the minimal system reduces to $N=9$ qubits, see Fig.~\ref{fig:mintcm}~(B).

Finally, we have to point out that our main interest lies in equivalent,
that is \textit{local unitary (LU) equivalent} graph states. However, computing LU-classes
of graph states turns out to be exceedingly difficult whereas equivalence
classes with respect to \textit{local Clifford (LC)} operations can be computed in principle
for any number of qubits algorithmically 
\cite{Bouchet1991,Bouchet199375,VandenNest2004,1_hein_EIGSAIA_S}, though not efficiently.
For graph states it is known \cite{1_hein_MPEIGS_A} that up to $7$ qubits LU-
and LC-classes coincide. Unfortunately, the conjecture that this equivalence holds
for arbitrary numbers of qubits (see \cite{1_werner_SOPQIT} and references therein)
was disproved by means of a $27$-qubit counterexample \cite{Ji2010}.
However, it is believed \cite{Ji2010} (but so far not proved) that this counterexample 
is the smallest one.
Since the systems considered in this letter are well below this threshold,
we will deal with LC- instead of LU-equivalent graph states in order to
simplify computations.


This paper is organized as follows. In the first chapter (\ref{sec:preliminaries})
we summarize the mathematical and physical prerequisites used throughout this work:
First, we outline the basics of the stabilizer formalism and give a short remark on
different notions of ``equivalent'' states. Subsequently, two important 
classes of stabilizer states, graph states and surface codes, are reviewed in more detail.
In the second chapter (\ref{sec:mapping}) we present our first result, proving that there
is a simple geometrical procedure for the transformation of surface codes into LC-equivalent
graph states. This procedure is illustraded by means of two examples.
In the last chapter (\ref{sec:minimal}) we present our second result: the smallest
toric code setups for which no local graph state exists.
Readers familiar with the basics of graph theory, graph states and surface codes
may skip the preliminaries and proceed with Chapter \ref{sec:mapping}.
Experimentalists primarily interested in the proposed non-local setups may omit
the second chapter, as it is only of technical interest for the setups in question.

\section{Preliminaries}
\label{sec:preliminaries}


The settings considered in this paper are composed of $N$ qubits (or \textit{spins}) 
and characterized by their coupling interactions. Thus the states are described by 
elements of the $2^N$-dimensional Hilbert space 
$\mathcal{H}^N:=\bigotimes_{i=1}^N\mathbb{C}^2_{i}$.
The Pauli matrices acting on the Hilbert space of the $i$th qubit are
denoted by $\sigma^x_i$, $\sigma^y_i$, $\sigma^z_i$
and, for the sake of consistency, $\sigma^0_i=\mathds{1}$.
As a basis of the 1-qubit Hilbert space $\mathcal{H}^1$ one chooses usually the 
\textit{computational basis} $\left\{\Ket{0}, \Ket{1}\right\}$ defined as eigenbasis 
of $\sigma^z$. Henceforth we shall use the convention $\sigma^z\Ket{0}=\Ket{0}$ 
and $\sigma^z\Ket{1}=-\Ket{1}$.

In this paper we consider two distinguished classes of $N$-qubit states 
known as \textit{surface codes} and \textit{graph states}. Both are examples for the much 
larger and intensively studied class of \textit{stabilizer states}. Thus we will start 
with a short survey of the \textit{stabilizer formalism}.


This formalism is widely used in quantum information theory and describes
a certain class of N-qubit states, called \textit{stabilizer states}, and their transformations
by means of \textit{stabilizer groups}.

To this end define the \textit{Pauli group} 
$G_1:=\operatorname{span}\left\{\sigma^x,\sigma^y,\sigma^z\right\}$
operating on a 1-qubit Hilbert space and generated by the Pauli matrices.
The generalization for $N$ qubits is defined as the $N$-fold tensor product 
$G_N:=\bigotimes_{i=1}^NG_1^{(i)}$ of the 1-qubit Pauli group.
Now let $\mathcal{G}=\left\{g_i\right\}_{i=1}^{d}\subseteq G_N\setminus\left\{-\mathds{1}\right\}$
be a given set of independent, pairwise commuting Pauli operators. 
Then $\mathcal{S}:=\operatorname{span}\mathcal{G}$
is called \textit{stabilizer} of the \textit{protected subspace}
$\mathcal{PS}=\left\{\Ket{\Phi}\in\mathcal{H}^N\,|\,\mathcal{S}\Ket{\Phi}=\Ket{\Phi}\right\}$.
$\mathcal{G}$ is the \textit{generating set} of $\mathcal{S}$ and 
$d=|\mathcal{G}|=:\operatorname{rank}\mathcal{S}$
is called \textit{rank} of the stabilizer. The dimension of the proteced space 
is $\dim\mathcal{PS}=2^{N-d}$ (for a proof and further reading see 
\cite{NielsenChuang201012}), thus there are $2^{N-d}$ independent \textit{stabilizer states} 
in $\mathcal{PS}$.

It proves advantageous to describe $N$-qubit states $\Ket{\Psi}\in\mathcal{PS}$
by means of their stabilizing operators generated by $\mathcal{G}$, for one has
to track at most $N$ generators whereas the description of general
$N$-qubit states requires an overall number of $2^N$ complex coordinates.
In order to describe unitary \textit{transformations} of stabilizer states by 
transformations of their stabilizer, one may use only unitaries that leave $G_N$ 
invariant under conjugation. In this context, one defines the \textit{Clifford group}
$C_1:=N_{U\left(2\right)}\left(G_1\right)=\left\{U\in U\left(2\right)\,|\,UG_1U^\dagger=G_1 \right\}$
as normalizer of the Pauli group in the $1$-qubit unitary group $U(2)$, 
and canonically extends this definition to the \textit{local Clifford group}
$C_N^l:=\bigotimes_{i=1}^N C_1^{(i)}$ operating on an $N$-qubit Hilbert space.
If the transformations of a given stabilizer state are restricted to $C_N^l$ 
(or more generally $C_N$, see
\footnote{The \textit{general Clifford group} is defined as 
  ${C}_N:=N_{U\left(2^N\right)}\left(G_N\right)=\left\{u\in U\left(2^N\right)\,|\,uG_Nu^\dagger=G_N \right\}$
  and includes \textit{non-local} unitary operations as well.}),
it is sufficient to transform the generators $\mathcal{G}\rightarrow C\mathcal{G}C^\dagger$
to describe the transformation $C\in C_N^l$ in the stabilizer formalism 
since $C\mathcal{G}C^\dagger$ once more meets the requirements of a generating set.

Whereas in the stabilizer framework local Clifford operations play a crucial role,
one generally deals with a larger class of local operations,
namely \textit{local unitaries} $U\in U_N^l:=\bigotimes_{i=1}^N U(2)^{(i)}$.
Since two arbitrary multi-qubit states cannot be transformed into one another
by local Clifford operations or even local unitaries in general, the class
of $N$-qubit states decomposes into equivalence classes with respect to
restricted sets of transformations.


To this end let $\Ket{\Psi}$ and $\Ket{\Psi^\prime}$ be two arbitrary $N$-qubit states.
They are called \textit{equivalent under local unitary operations} 
(\textit{LU-equivalent} or \textit{equivalent} for short) 
if there exists $U\in U_N^l$, such that $\Ket{\Psi}=U\Ket{\Psi^\prime}$;
write $\Ket{\Psi}\sim_{\mathrm{LU}}\Ket{\Psi^\prime}$.
Accordingly one calls them \textit{equivalent under local Clifford operations} 
(\textit{LC-equivalent}) 
if there exists $C\in C_N^l$, such that $\Ket{\Psi}=C\Ket{\Psi^\prime}$;
write $\Ket{\Psi}\sim_{\mathrm{LC}}\Ket{\Psi^\prime}$.

Since $C_N^l$ is a proper subgroup of $U_N^l$, it follows $\mathrm{LC}\Rightarrow\mathrm{LU}$ but
as $C_N^l\neq U_N^l$ the converse cannot be true in general.
However, due to the lack of counterexamples it was conjectured 
\cite{1_werner_SOPQIT} that for the smaller class of \textit{stabilizer states} 
LU- and LC-equivalence coincide: $\mathrm{LC}\Leftrightarrow\mathrm{LU}$.
Although it was shown that this holds for a large class of 
stabilizer states \cite{vandennest2005,zeng2007} the LU-LC-conjecture was disproved 
by Ji \textit{et al.} in 2009 \cite{Ji2010}.
Nevertheless it was shown in Ref. \cite{1_hein_MPEIGS_A} 
that $\mathrm{LU}\Leftrightarrow\mathrm{LC}$ 
holds for graph states with up to $7$ qubits. Furthermore it is believed \cite{Ji2010} 
(but so far not proved) that the counterexample used in Ref. \cite{Ji2010} is the 
smallest one ($N=27$). Recently the class of stabilizer states for which the LU-LC-conjecture 
holds was enlarged \cite{Sarvepalli2010}, including a large class of surface codes.

After this brief survey of basic concepts of the stabilizer formalism and the notion of 
LC-equivalence we may now define a special class of stabilizer states, 
called \textit{graph states}.


Graph states were introduced by Hein \textit{et al.} \cite{1_hein_MPEIGS_A} in 2004
and denote a class of $N$-qubit states that can be described by \textit{simple graphs}.
In this context it is shown \cite{1_hein_MPEIGS_A} 
that the aforementioned LC-equivalence is closely
related to the concept of \textit{local complementations} known in graph theory.
To state the concepts of graph states more precisely, basics of graph theory are required.
Thus we start with a short review of the mathematics that may be skipped by the reader familiar
with the concepts of graph theory.


A \textit{finite undirected graph} is a pair $G=(V,E)$ where $V$ denotes a finite set
of \textit{vertices} and $E\subseteq \left\{\{a,b\}\,|\,a,b\in V\right\}$ a set of \textit{edges}
connecting these vertices.
If there are neither loops (i.e. edges connecting a vertex to itself) nor multiple edges
(i.e. vertices connected by more than just one edge) $G$ is called a \textit{simple graph}. 
In contrast, if there are multiple edges (i.e. $E$ is a multiset) but no loops \footnote{
Some authors include the possibility of loops in their defintion of multigraphs.} $G$ is called
a \textit{multigraph}. In the following, simple graphs shall be denoted by 
upper-case letters $G,H,\dots$ and multigraphs by calligraphic 
letters $\mathcal{L},\mathcal{M},\dots$.

For a given graph $G=(V,E)$ the \textit{functions} $\mathbb{V}(G)$ and $\mathbb{E}(G)$ 
denote the sets of vertices and edges, i.e. $\mathbb{V}(G)=V$ and $\mathbb{E}(G)=E$. 
The \textit{order} of a graph $G=(V,E)$
is the number of its vertices and denoted by $\left|G\right|:=\left|V\right|$.
One naturally defines the \textit{neighbourhood} of a vertex $v\in V$ as the
subset $N_v:=\left\{w\in V\,|\,\{w,v\}\in E\right\}\subseteq V$ of vertices directly
connected to $v$. For the sake of brevity we shall denote the \textit{complete graph}
over a vertex set $V$ by 
$\left<V\right>:=\left(V,\left\{\{v,w\}\,|\,v,w\in V;\,v\neq w\right\}\right)$,
that is, the unique simple graph with edges connecting any pair of vertices.

Graph theory deals not only with properties of graphs but also with transformations of the latter.
A particularly interesting transformation in the context of graph states is 
known as \textit{local complementation}.
The \textit{local complement} $\tau_v(G)$ of a simple graph $G=(V,E)$ at vertex $v\in V$ 
is formally defined as $\tau_v(G):=G+\left<N_v\right>$, where $+$ denotes the binary 
addition of edges.
Practically the procedure described by $\tau_v$ reads as follows. 
Focussing exclusively on the neighbours of $v$ one deletes all edges in $E$ 
connecting these vertices and adds new edges between formerly unconnected
neighbours. This yields $\tau_v(G)$.

In conclusion we note that two graphs $G=(V,E)$ and $H=(V,F)$ over the same vertex 
set $V$ are called \textit{equivalent under local complementations} (\textit{LC-equivalent}) 
if and only if there is a sequence of local complementations 
$\tau=\tau_{v_1}\circ\tau_{v_2}\circ\dots\circ\tau_{v_N}$ such that
$G=\tau(H)$; write $G\sim_{\mathrm{LC}} H$.
In the context of graph states, overloading ``LC-equivalent'' proves convenient and is 
therefore intended; this will become clear below.


We may now give the formal definition of graph states and comment on their properties.
To this end let $G$ be an arbitrary simple graph with edges $E$ and vertices $V$. Consider
a system of $N=\left|G\right|$ qubits -- each identified with a \textit{vertex} $v\in V$ --
and define the (independent) operators 
\begin{equation}
  K_G^{v}:=\sigma_v^x\prod\limits_{w\in N_v}\sigma_{w}^z\in G_N\,.
\end{equation}
Then $\mathcal{G}[G]=\left\{K_G^{v}\,|\,v\in V\right\}$ is a generating set
of the \textit{graph state stabilizer} $\mathcal{S}[G]=\operatorname{span}\mathcal{G}[G]$.
If there is no danger of confusion, we shall omit the dependence on $G$, 
viz. $\mathcal{S}\equiv\mathcal{S}[G]$.
Since there are $N$ independent generators, the protected space $\mathcal{PS}[G]$ 
is one-dimensional. The state stabilized by $\mathcal{S}[G]$, henceforth characterized 
by $G$, is called \textit{graph state} and denoted by $\Ket{G}$.

In this paper we shall define graph states in this stabilizer representation.
However, there is an alternative representation which proposes a simple procedure
for the construction of graph states in terms
of controlled $z$-gates 
$U_{ab}=\mathcal{P}^{z,+}_{a}\otimes\mathds{1}_b+\mathcal{P}^{z,-}_{a}\otimes\sigma_b^z$.
Here $\mathcal{P}^{z,\pm}_{v}=\frac{1}{2}\left[\mathds{1}_v\pm\sigma^z_v\right]$ denote
the projectors operating on qubit $v$.
Then it can be shown \cite{1_hein_MPEIGS_A} that
\begin{equation}
  \label{eq:graphstate2}
  \Ket{G}=\prod\limits_{\{a,b\}\in \mathbb{E}(G)}U_{ab}\Ket{+}^{\mathbb{V}(G)}
\end{equation}
where $\Ket{+}=1/\sqrt{2}\left(\Ket{0}+\Ket{1}\right)$ and $\Ket{+}^{\mathbb{V}(G)}$ 
denotes the $N$-fold tensor product of states indexed by the vertices $\mathbb{V}(G)$.
According to Eq. (\ref{eq:graphstate2}) the empty graph on $N$ vertices $V$
is represented by the product-state $\Ket{+}^{V}$.
One may add an edge between two vertices $a,b\in V$ by applying a controlled $z$-gate $U_{ab}$
on the corresponding qubits. This yields a handy procedure for the construction of 
any graph state $\Ket{G}$ on $N$ qubits -- edge by edge -- provided an entropy 
free initial state $\Ket{+}^{\mathbb{V}(G)}$.

Apart from their implementation,
graph states are particularly interesting multi-qubit states since several physical properties
can be related to graph theoretic properties of the underlying graph \cite{1_hein_MPEIGS_A}. 
One of these properties is the aforementioned local Clifford equivalence:
Let $\Ket{G}$ and $\Ket{H}$ be two graph states with corresponding graphs 
$G=(E,V)$ and $H=(F,V)$.
Then they are LC-equivalent if and only if their graphs are connected by a
sequence of local complementations. Formally this reads
$\Ket{G}\sim_{\mathrm{LC}}\Ket{H}\Leftrightarrow G\sim_{\mathrm{LC}} H$.
The notion of LC-equivalent \textit{graphs} was known in graph theory for a while
and it was shown that there is an efficient algorithm to check whether two graphs
are LC-equivalent \cite{Bouchet1991,Bouchet199375}.
Thus the aforementioned relation between the LC-equivalence of \textit{graph states} 
and \textit{graphs} provides an efficient algorithm to decide whether two graph states 
belong to the same LC-class. 

For this reason the question arises whether this insight into the nature of LC-equivalence 
may be extended to a larger class of multi-qubit states. Indeed, it was 
shown \cite{VandenNest2004,Schlingemann2002}
that graph states are universal standard forms of stabilizer states. That is to say,
for any $N$-qubit stabilizer state $\Ket{\Psi}$ there exists an LC-equivalent 
graph state $\Ket{G_{\Psi}}$. Since the transformation 
$\Ket{\Psi}\stackrel{\mathrm{LC}}{\longrightarrow}\Ket{G_{\Psi}}$ is efficiently computable,
the problem of LC-equivalence for stabilizer states is reduced to the class of graph states --
where it can be solved by means of mere graph theory.

Besides graph states the stabilizer formalism is widely used for the description
of quantum codes. A prominent example we are particularly interested in
is the \textit{toric code model} (tcm) and its generalization, the \textit{surface codes}.


In 2002 Kitaev proposed the qubit model named \textit{toric code} \cite{1_kitaev_FTQCBA_A} 
showing the inherent capability of quantum error correction on a physical level and 
a topologically ordered phase at zero temperature. Whereas the tcm is typically 
defined on a square lattice embedded into a 2-torus, one may generalize this idea
to arbitrary (multi-) graphs on orientable compact manifolds, called
\textit{surface codes} \cite{Sarvepalli2010,Bravyi1998,1_kitaev_FTQCBA_A}.

\begin{figure}[tbp]
\includegraphics[width=0.9\linewidth]{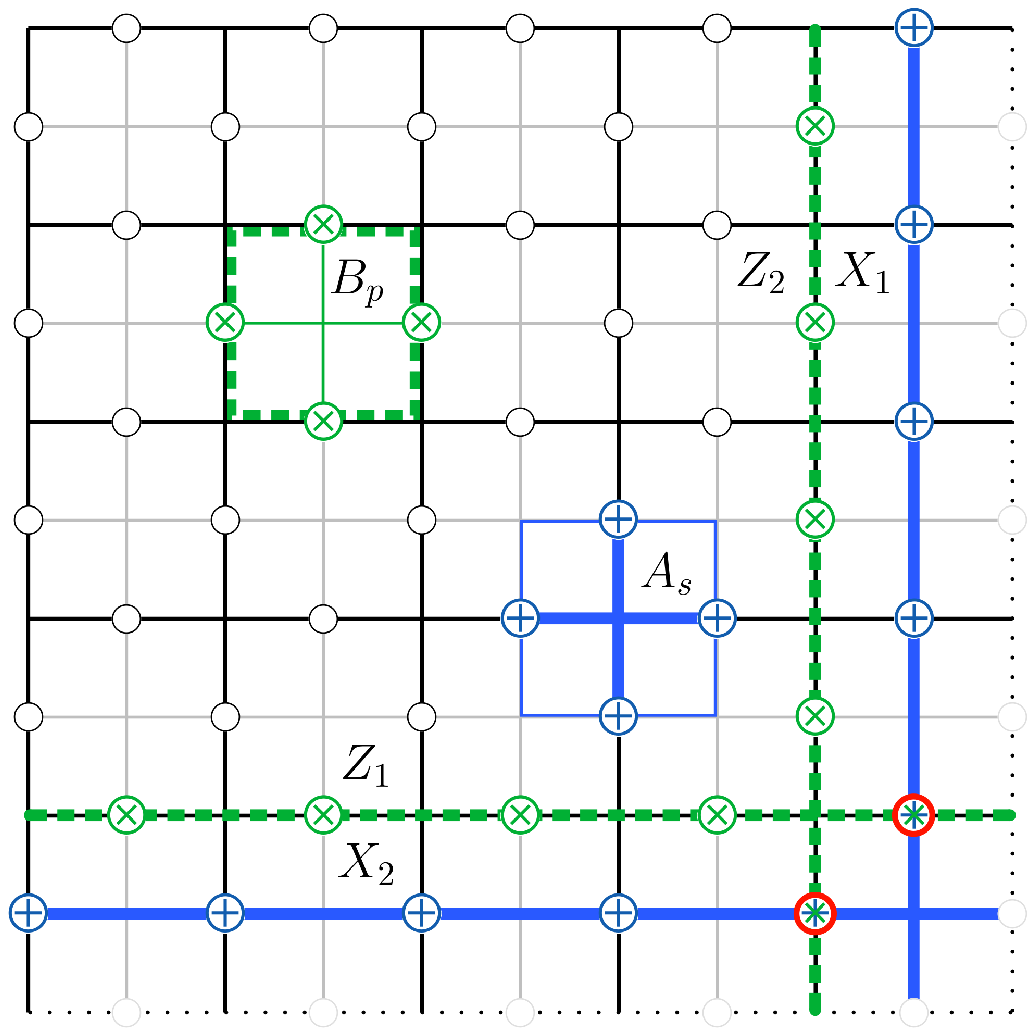}
\caption{\label{fig:tcm_1}(Color online) 
  The tcm with qubits 
  (\protect\raisebox{-0.7mm}{\protect\hspace{-0.2mm}\protect\includegraphics[scale=0.35]{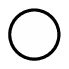}}) 
  located on the edges of a square lattice embedded into a $2$-torus. 
  The $\sigma^x$ and $\sigma^z$ Pauli operators are denoted by
  \protect\raisebox{0.1mm}{\protect\hspace{-0.4mm}\protect\includegraphics[scale=0.35]{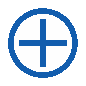}}
  and 
  \protect\raisebox{0.1mm}{\protect\hspace{-0.4mm}\protect\includegraphics[scale=0.35]{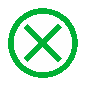}},
  respectively.
  The intersections of X- and Z-type loop operators, responsible for their commutation relations, 
  are highlighted by
  \protect\raisebox{0.1mm}{\protect\hspace{-0.0mm}\protect\includegraphics[scale=0.35]{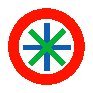}}.
  The upper [left] border is identified with the dashed lower [right] border.
}
\end{figure}

Aiming at the definition of surface codes, we need some notation regarding graph embeddings.
To this end let $\mathcal{L}$ be a multigraph and $\Sigma$ a closed, orientable 2-manifold.
A \textit{graph embedding} $\mathfrak{X}\equiv\mathfrak{X}_{\mathcal{L},\Sigma}$
of $\mathcal{L}$ into $\Sigma$ can be informally understood as a ``drawing'' of
$\mathcal{L}$ on the surface $\Sigma$ without intersecting edges.
In the following ``embedding'' has always to be read as \textit{2-cell embedding}, 
that is, a drawing on the surface, such that every face is homeomorphic to an open 
disc in $\mathbb{R}^2$ \footnote{More concrete: Provided an embedding of a graph into a surface,
  then ``deleting'' the graph decomposes the surface into several pieces (the \textit{faces} 
  of this embedding). If each of these pieces can be deformed continuously into a disc, the 
  embedding is called a \textit{2-cell embedding}. For instance, the common square-shaped 
  plaquettes of a toric code system are homeomorphic to open discs; therefore the toric code 
  is based on a 2-cell embedding of the square lattice into the 2-torus.}.
The \textit{genus} $g[\Sigma]$ of $\Sigma$
is its number of ``holes'' or ``handles''. The genus $g[\mathcal{L}]$ \textit{of a graph} is
defined as the least genus of surfaces, such that an embedding exists.
Each embedding $\mathfrak{X}_{\mathcal{L},\Sigma}$ induces a set of \textit{faces} denoted by
$\mathbb{P}_{\mathfrak{X}}(\mathcal{L})$ or simply $\mathbb{P}(\mathcal{L})$.
Given a graph embedding $\mathfrak{X}_{\mathcal{L},\Sigma}$
the \textit{dual graph} $\mathcal{L}^\ast=(V^\ast,E^\ast)$ is defined so that
$V^\ast:=\mathbb{P}_{\mathfrak{X}}(\mathcal{L})$
and two dual vertices $v_1^\ast,v_2^\ast\in V^\ast$ are adjacent, 
i.e. $\{v_1^\ast,v_2^\ast\}\in E^\ast$, iff there is a common edge in 
$\mathbb{E}(\mathcal{L})$ to $v_1^\ast$ and $v_2^\ast$ as faces. A \textit{(co-)cycle}
$\mathcal{C}\subseteq\mathbb{E}(\mathcal{L})$ 
($\mathcal{C}^\ast\subseteq\mathbb{E}(\mathcal{L}^\ast)$)
on $\mathcal{L}$ ($\mathcal{L}^\ast$) is a set of edges, such that each vertex is 
adjacent to an even number of these edges. Are there at most two of these edges joining 
at each vertex and is $\mathcal{C}$ ($\mathcal{C}^\ast$) connected within the graph,
then it is called a \textit{simple (co-)cycle}, namely, a single cycle without self-intersections.

For the definition of surface codes,
consider a multigraph $\mathcal{L}$ 2-cell embedded via $\mathfrak{X}_{\mathcal{L},\Sigma}$ 
into a surface $\Sigma$ of genus $g[\Sigma]\geq g[\mathcal{L}]$ and a system 
of $\left|\mathbb{E}(\mathcal{L})\right|$ qubits identified with the \textit{edges} 
of $\mathcal{L}$ (cf. graph states!). Define the \textit{star} and \textit{plaquette operators}
\begin{equation}
  A_s:=\bigotimes\limits_{i\in s}\sigma_i^x\qquad\text{and}\qquad
  B_p:=\bigotimes\limits_{i\in p}\sigma_i^z
\end{equation}
for each vertex $s\in \mathbb{V}(\mathcal{L})$ and face 
$p\in \mathbb{P}_{\mathfrak{X}}(\mathcal{L})$.
Here we identify a \textit{face} or \textit{plaquette} 
$p\in \mathbb{P}_{\mathfrak{X}}(\mathcal{L})$
with its set of \textit{peripheral edges} and a \textit{vertex}
$v\in\mathbb{V}(\mathcal{L})$ with its \textit{adjacent edges}.
Then $\mathcal{S}\equiv\mathcal{S}[\mathfrak{X}]:=\operatorname{span}
\left\{A_s,B_p\,|\,s\in \mathbb{V}(\mathcal{L}),\,p\in \mathbb{P}_{\mathfrak{X}}(\mathcal{L})\right\}$
is the stabilizer of the \textit{surface code}
corresponding to the chosen embedding $\mathfrak{X}_{\mathcal{L},\Sigma}$.
Its degeneracy is $\dim\mathcal{PS}=4^g=2^{2g}$ in the case of closed boundary 
conditions \cite{Bravyi1998} as one can show easily using \textit{Eulers formula} for 
2-cell embeddings \footnote{If $\mathcal{L}$ is a multigraph of genus $g[\mathcal{L}]$ 
  2-cell embedded into an orientable compact surface $\Sigma$
  of genus $g\equiv g\left[\Sigma\right]\geq g[\mathcal{L}]$, it holds $v+f-e=2-2g$
  where $v$, $e$ and $f$ denote the number of vertices, edges and faces, respectively.
}.
A possible basis of the protected space $\mathcal{PS}$ is defined and manipulated by 
means of \textit{loop operators}
\begin{equation}
  Z_k\equiv Z\left[\mathcal{C}_{k}\right]:=\bigotimes\limits_{i\in \mathcal{C}_{k}}\sigma_i^z
  ,\quad
  X_k\equiv X\left[\mathcal{C}^\ast_{k}\right]:=\bigotimes\limits_{i^\ast\in 
    \mathcal{C}^\ast_{k}}\sigma_{i^\ast}^x
\end{equation}
where the simple (co-)cycles $\mathcal{C}_{k}\subseteq \mathcal{L}$ and 
$\mathcal{C}^\ast_{k}\subseteq \mathcal{L}^\ast$ are chosen homologously distinct for 
different $k$ and non-trivial (see Fig. \ref{fig:tcm_1}).
In algebraic topology, two paths on a compact, orientable 2-manifold $\Sigma$ are 
considered \textit{homologous} if their union is the boundary of a surface embedded 
into $\Sigma$. In our case, two cycles on the embedded graph $\mathcal{L}$
are homologous iff their union is the boundary of a union of faces. This defines an 
equivalence relation $\sim$ on the group of all cycles $\mathfrak{C}\left(\mathcal{L}\right)$.
The quotient $H_1:=\bigslant{\mathfrak{C}\left(\mathcal{L}\right)}{\sim}$ is 
called \textit{first homology group} and contains homology classes $[\mathcal{C}]$
of cycles $\mathcal{C}\in\mathfrak{C}\left(\mathcal{L}\right)$.
Restricted to the protected space $\mathcal{PS}$, the effect of a loop 
operators $Z\left[\mathcal{C}\right]$ ($X\left[\mathcal{C}^\ast\right]$) depends 
\textit{solely} on the homology class of $\mathcal{C}$ ($\mathcal{C^\ast}$)
since they act trivially if the (co-)cycle is homologeously trivial, that is, 
$\mathcal{C}$ ($\mathcal{C}^\ast$) is the boundary of a union of faces (stars, i.e. dual faces). 
Therefore, by a slight abuse of notation, we may regard $Z\left[\mathcal{C}\right]$
either as a mapping $Z[\,\cdot\,]$ with domain $\mathfrak{C}\left(\mathcal{L}\right)$, 
or as an operator $Z\cdot$ with domain $H_1$.

If the indices $k$ are chosen properly, the loop operators fulfil the algebra
$\left\{Z_k,X_k\right\}=0=\left[Z_k,X_l\right]$ for $k\neq l$.
Then $\left\{\Ket{\vec v}\in\mathcal{PS}\,|\,
Z_k\Ket{\vec v}=v_k\Ket{\vec v},\,1\leq k\leq 2g\right\}$
is a basis of $\mathcal{PS}$, where the components of $\vec v\in\{-1,1\}^{2g}$ are called
\textit{topological quantum numbers} and represent the \textit{logical} qubits encoded 
in the protected space. The topological quantum numbers are flipped by the $2g$ distinct $X_k$
loop operators via $X_k\Ket{v_1,\dots,v_k,\dots,v_{2g}}=\Ket{v_1,\dots,-v_k,\dots,v_{2g}}$
due to the aforementioned (anti-)commutation relations.

To conclude this section, we note that in the special case of a
square lattice $\mathcal{L}$ and $\Sigma=\mathbb{T}^2$ a $2$-torus, 
as depicted in Fig.~\ref{fig:tcm_1}, this construction yields the tcm \cite{1_kitaev_FTQCBA_A}.

\section{Mapping surface codes to graph states}
\label{sec:mapping}

\subsection{Results}

In the following, let $\Ket{\vec v}$ be the state of a surface code with topological 
quantum numbers $\vec v$, corresponding stabilizer $\mathcal{S[\mathfrak{X}]}$ and 
embedded multigraph $\mathcal{L}$. $\Ket{G}$ denotes a graph state with simple 
graph $G$ and stabilizer $\mathcal{S}[G]$.
Since every stabilizer state is LC-equivalent to some graph state, there exists a 
local Clifford operation $C\in\mathcal{C}^l_N$ such that $\Ket{G}=C\Ket{\vec v}$.
We are now interested in finding $C$ and $\mathcal{S}[G]$ for a given surface 
code $\mathcal{S[\mathfrak{X}]}$ in state $\Ket{\vec v}$.
The binary algorithm developed by Van den Nest \textit{et al.} \cite{VandenNest2004}
allows one to perform this computation algorithmically. Here we show that in the case
of surface codes there is a purely \textit{geometrical}, graph theoretic transformation rule.

\begin{figure}[tbp]
  \includegraphics[scale=0.39]{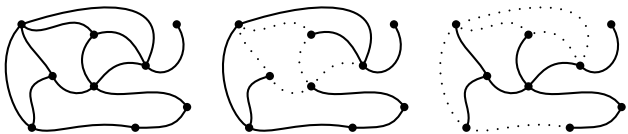}  
  \caption{\label{fig:tree}
    The graph on the left hand side is simple and \textit{not} a tree since
    it features cycles. Deleting edges that support a cycle (illustrated
    by dotted lines) yields two non-isomorphic spanning trees.}
\end{figure}

To this end a further notion of graph theory is required.
Let $\mathcal{L}$ and $\mathcal{L}^\prime$ be (multi-)graphs.
$\mathcal{L}^\prime$ is called a \textit{spanning tree} of $\mathcal{L}$ if it is simple, 
connected, devoid of cycles and 
$\mathbb{V}\left[\mathcal{L}\right]=\mathbb{V}\left[\mathcal{L}^\prime\right]$
as well as $\mathbb{E}\left[\mathcal{L}^\prime\right]\subseteq\mathbb{E}\left[\mathcal{L}\right]$.
That is to say, spanning trees are distinguished subgraphs without cycles that cover all 
vertices of the original graph. However, they are \textit{not unique} in general.
Spanning trees of a given graph may be constructed by deleting a set 
$\mathcal{E}\subseteq \mathbb{E}[\mathcal{L}]$ of edges such that there is no cycle 
left and the graph remains connected.
For an example of this procedure see Fig. \ref{fig:tree}.
Let us now introduce a mapping that converts multigraphs to a particular class of simple graphs
which proves crucial for the LC-transformation of surface codes to graph states.
\begin{Definition}[$\varphi_{\mathcal{L}^\prime}\,:\,\mathcal{L}\mapsto G$]
  \label{def:map}
  Let $\mathcal{L}$ be a multigraph and $\mathcal{L}^\prime$ one of its spanning trees.
  Then $G=\varphi_{\mathcal{L}^\prime}(\mathcal{L})$ is a simple graph defined by the 
  following construction:

  Set $\mathbb{V}(G):=\mathbb{E}(\mathcal{L})$ as the vertex set of $G$.
  For $r,s\in \mathbb{V}(G)$ the set $\{r,s\}$ is an edge of $G$ if and only if
  there is a path $\mathcal{C}_{pq}$ from $p$ to $q$ on 
  $\mathcal{L}^\prime$ ($p,q\in \mathbb{V}\left(\mathcal{L}^\prime\right)=\mathbb{V}\left(\mathcal{L}\right)$) 
  such that $r=\{p,q\}$ and $s\in\mathcal{C}_{pq}$.
\end{Definition}

The physical relevance of this mapping from multigraphs 
to simple graphs is given by the following

\begin{Theorem}[LC-equivalence]
  \label{thm:1}
  Given the surface code state $\Ket{\mathcal{\vec v}}$ on $\mathcal{L}$
  and any spanning tree $\mathcal{L}^\prime$.
  Then the graph state $\Ket{G}=\Ket{\varphi_{\mathcal{L}^\prime}(\mathcal{L})}\in\mathcal{H}^N$ 
  is LC-equivalent to $\Ket{\vec v}$.
\end{Theorem}

As a final remark note that the combination of the LC-rule for 
graph states \cite{1_hein_MPEIGS_A} and 
Theorem \ref{thm:1} yields the purely graph theoretic statement that for arbitrary 
spanning trees $\mathcal{L}^\prime_1$ and $\mathcal{L}^\prime_2$ it holds 
$\varphi_{\mathcal{L}^\prime_1}(\mathcal{L})\sim_{\mathrm{LC}}\varphi_{\mathcal{L}^\prime_2}(\mathcal{L})$.
For the sake of consistency, we derived a direct graph theoretic proof
for this very statement (which shall be omitted here).

\subsection{Examples}
\label{subsec:examples}

\begin{figure}[tbp]
  \small
  A\includegraphics[width=0.59\linewidth]{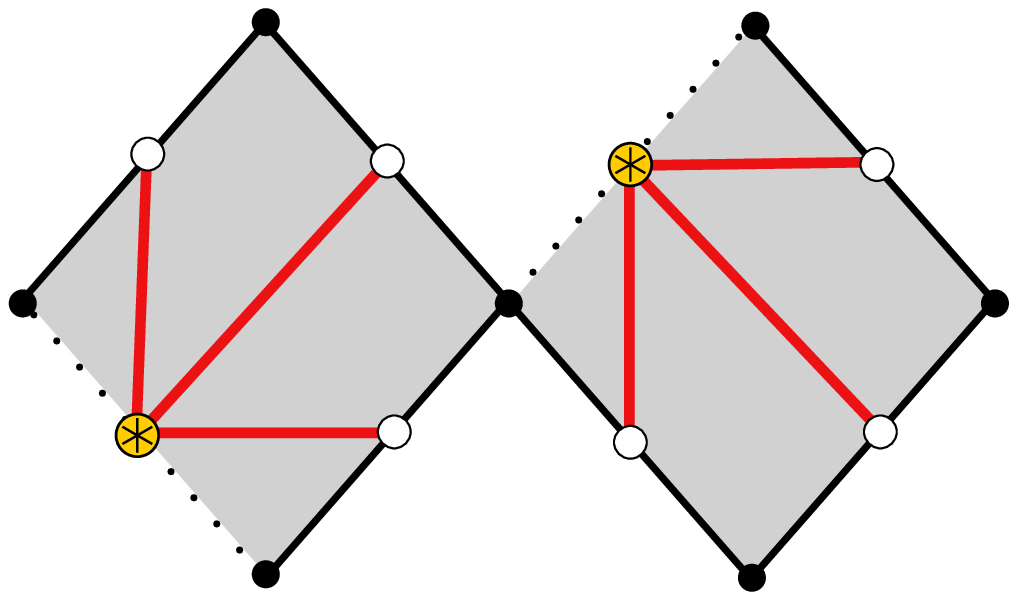}
  B\includegraphics[width=0.29\linewidth]{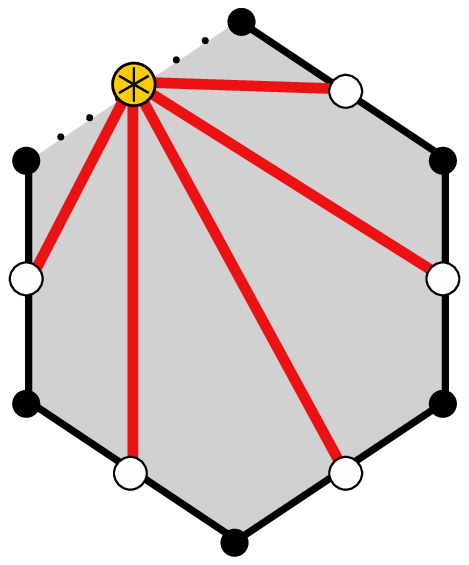}
  \normalfont
  \caption{\label{fig:example}(Color online) 
    Two simple examples of surface codes and LC-equivalent graph states.
    Product states (\textbf{A}) are mapped to disjoint graphs and GHZ states (\textbf{B}) 
    are transformed to star graphs. A description of the graphical notation is given in  
    Subsection \ref{subsec:examples}.}
\end{figure}

Here we discuss two examples, depicted in Fig. \ref{fig:example}, in order to demonstrate 
the construction implied by Def. \ref{def:map} and show the consistency with well known facts.
The graphical notation is as follows: 
Qubits are denoted by \raisebox{-0.7mm}{\hspace{-0.8mm}\includegraphics[scale=0.33]{qubit}}
whereas \raisebox{-0.7mm}{\hspace{-0.8mm}\includegraphics[scale=0.45]{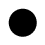}} denotes 
the vertices of the surface code, and thus the position of star operators $A_s$.
The edges of the surface code are denoted by thin, continuous and dashed lines connecting 
the vertices; the ``deleted'' edges $\mathcal{E}$ are drawn with dashed, the spanning 
tree with continuous lines. Faces are highlighted grey and symbolize plaquette operators $B_p$. 
Hadamard gates transforming qubits on ``deleted'' edges (see proof of Theorem \ref{thm:1} 
in subsection \ref{subsec:proof}) are symbolized by \infig{-0.2}{-0.6}{0.08}{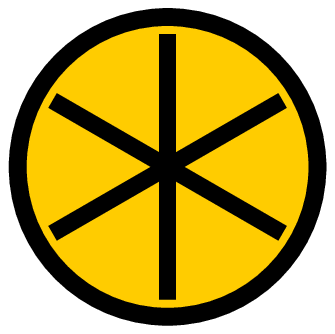}.
LC-equivalent graph states are drawn with bold red edges connecting the qubits.

\paragraph*{Example 1 (Product states).}

It is easy to see that protected states of two \textit{one-point connected} surface code systems
(Fig.~\ref{fig:example}~(A)) are product states, symbolically
$\left|\infig{-0.5}{0}{0.028}{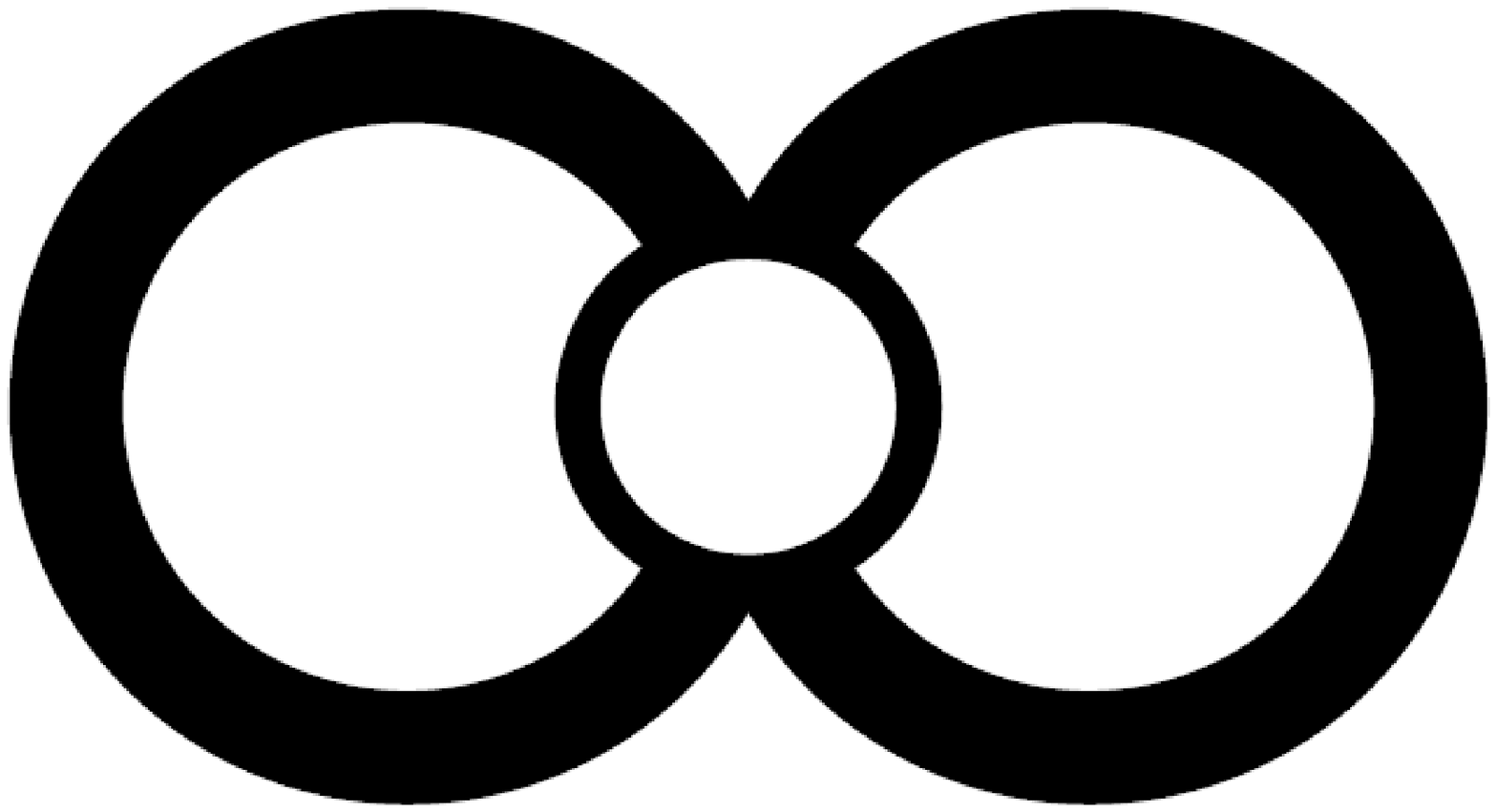}\right>=\left|\infig{-0.5}{0}{0.028}{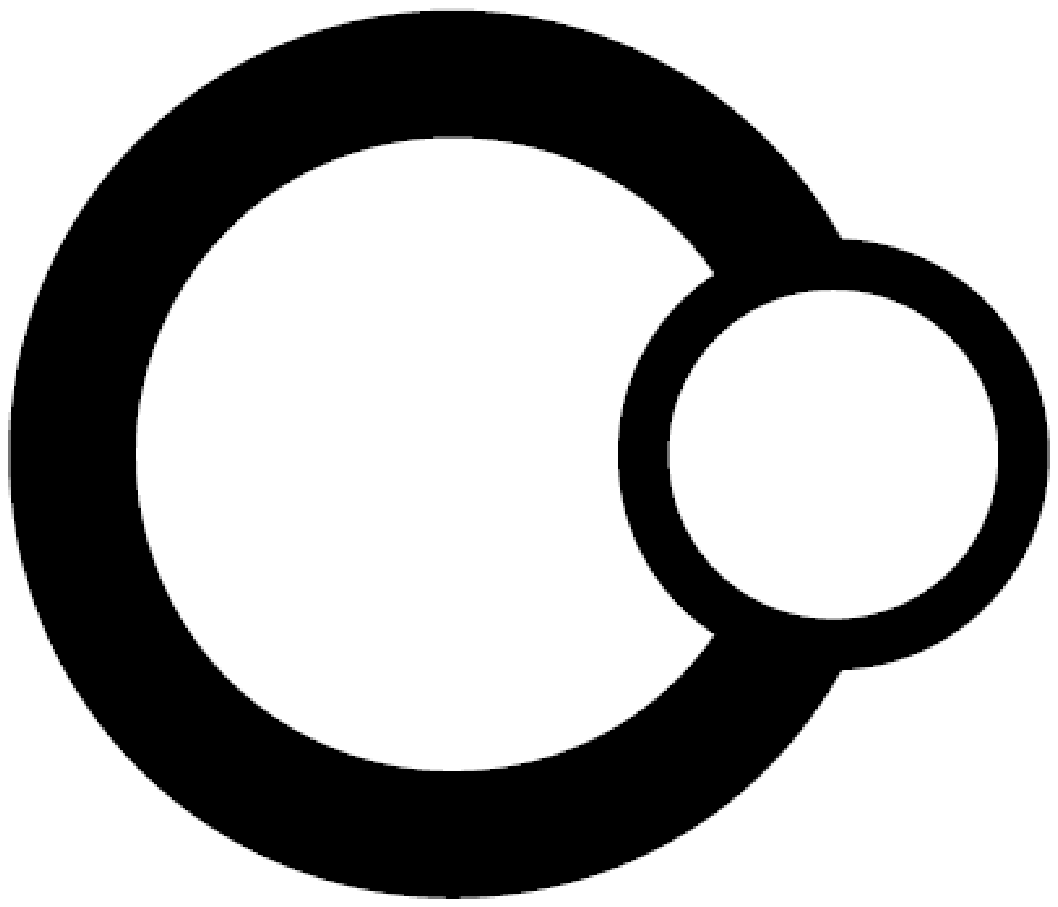}\right>\otimes\left|\infig{-0.5}{0}{0.028}{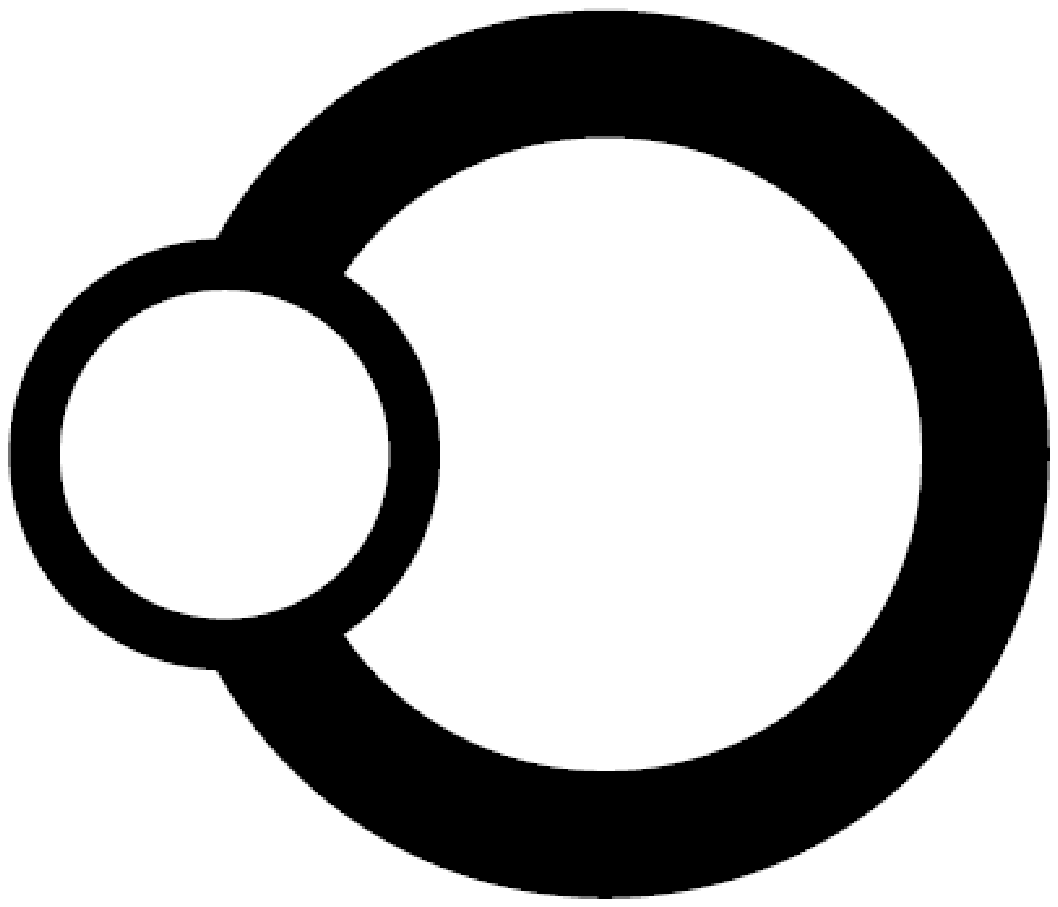}\right>$.
As a consequence of the one-point connection, any LC-equivalent graph state
turns out to be disconnected -- independent of the spanning tree used for the construction.
Since disconnected graphs denote the tensor product of graph states~\cite{1_hein_EIGSAIA_S}, 
we confirmed 
the statement $\left|\infig{-0.5}{0}{0.028}{fig1}\right>=\left|\infig{-0.5}{0}{0.028}{fig2}\right>\otimes\left|\infig{-0.5}{0}{0.028}{fig3}\right>$
in a purely geometric way, for local unitaries cannot alter the degree of entanglement. 

\paragraph*{Example 2 (GHZ states).}

One can show easily that the protected state of a single plaquette bounded by $N$ 
qubits is LC-equivalent to the $N$-qubit GHZ state $\Ket{\mathrm{GHZ}_N}$ since it has the form
$\left|\infig{-0.5}{0}{0.028}{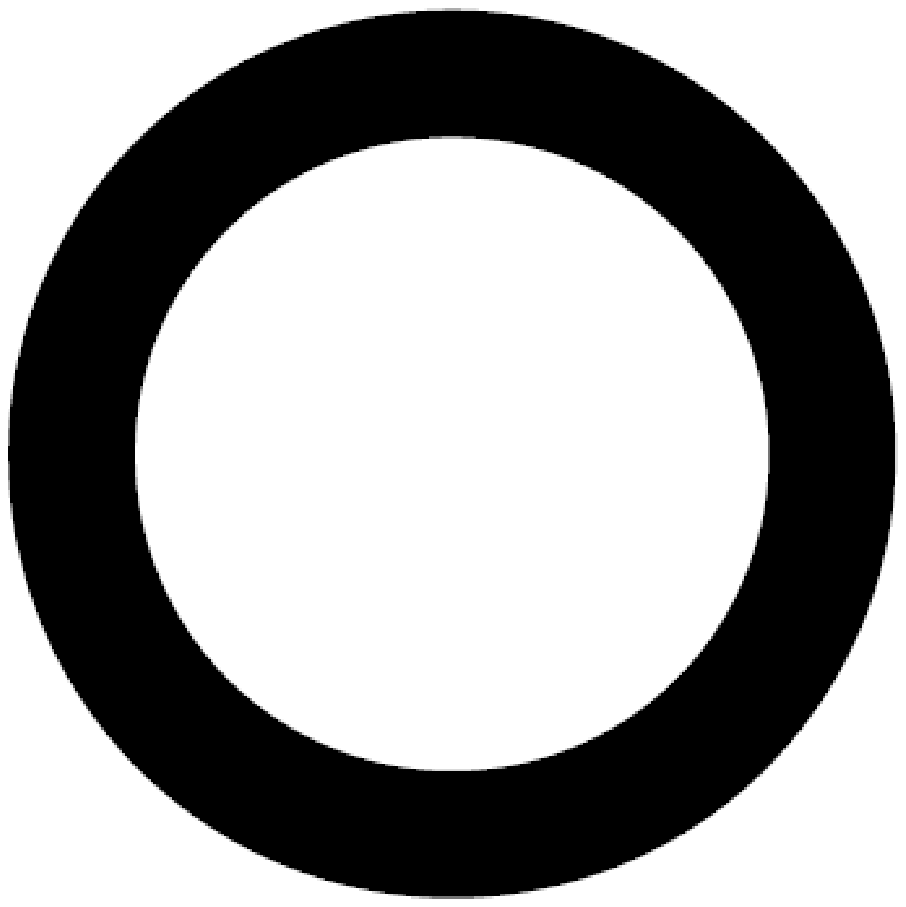}\right>=1/\sqrt{2}\left(\Ket{+}^{\otimes N}+\Ket{-}^{\otimes N}\right)$.
Consider the hexagonal plaquette depicted in Fig.~\ref{fig:example}~(B).
Constructing spanning trees is easy since ``deleting'' an arbitrary edge yields one.
Consequently any graph state constructed this way must be a \textit{star graph}.
Indeed, in \cite{1_hein_EIGSAIA_S} it is shown that star graphs (and complete graphs) are 
LC-equivalent to GHZ states.

\subsection{Comments on the structure of $\varphi_{\mathcal{L}^\prime}(\mathcal{L})$}
\label{subsec:comments}

The geometrically constructed graph $\varphi_{\mathcal{L}^\prime}(\mathcal{L})$
exhibits an interesting graph theoretical structure, which in turn furnishes insight into
the entanglement structure of the corresponding surface code state. 
To this end, we introduce the \textit{cycle space} $\mathfrak{C}\left(\mathcal{L}\right)$ and 
its dual, the \textit{cut space} $\mathfrak{D}\left(\mathcal{L}\right)$,
as linear subspaces of the edge space $\mathfrak{E}\left(\mathcal{L}\right)$; the latter 
being the power set of edges $\mathbb{E}\left(\mathcal{L}\right)$ equipped with binary
addition and multiplication. 
The cycle space $\mathfrak{C}\left(\mathcal{L}\right)$ is defined as set of all cycles 
in $\mathcal{L}$ (as sets of edges) equipped with the binary structure inherited by
$\mathfrak{E}\left(\mathcal{L}\right)$.
A \textit{cut} of $\mathcal{L}$ is a bipartition of its vertices;
the \textit{cut-set} of a given cut is the set of edges in $\mathcal{L}$ connecting these two disjoint sets 
of vertices (often the cut-set is just called \textit{cut}). The cut space $\mathfrak{D}\left(\mathcal{L}\right)$
is then defined as set of all cut-sets on $\mathcal{L}$ equipped with the usual binary structure.
Bases of both subspaces may be constructed by means of an 
arbitrary spanning tree $\mathcal{L}^\prime$ as follows. For each edge 
$e=\{p,q\}\in\mathbb{E}\left(\mathcal{L}\right)\setminus \mathbb{E}\left(\mathcal{L}^\prime\right)$
choose the cycle $\mathcal{C}_e:=\mathcal{C}_{pq}+e$ where $\mathcal{C}_{pq}$ denotes the 
unique path on the spanning tree connecting $p$ and $q$.
$\mathcal{C}_e$ is called \textit{fundamental cycle} on $\mathcal{L}$ with respect to 
$\mathcal{L}^\prime$ and one can show easily \cite{Diestel201010} that the set of all fundamental cycles 
$\left\{\mathcal{C}_e\,|\,e\in\mathbb{E}\left(\mathcal{L}\right)\setminus \mathbb{E}\left(\mathcal{L}^\prime\right)\right\}$
is a basis of the cycle space. Conversely, each edge 
$e^\prime\in\mathbb{E}\left(\mathcal{L}^\prime\right)$ induces a bipartition 
$\mathbb{V}_r^{e^\prime}\,\dot\cup\,\mathbb{V}_b^{e^\prime}=\mathbb{V}\left(\mathcal{L}\right)$ of the vertices, according to 
the component of the spanning tree they belong to if $e^\prime$ is deleted. 
The corresponding cut-set $\mathcal{D}_{e^\prime}$ is called \textit{fundamental cut} 
of $e^\prime$ with respect to $\mathcal{L}^\prime$. Analogously the set of all fundamental cuts
$\left\{\mathcal{D}_{e^\prime}\,|\,e^\prime\in\mathbb{E}\left(\mathcal{L}^\prime\right)\right\}$ 
yields a basis of the cut space.

\begin{figure}[b]
  \includegraphics[width=1.0\linewidth]{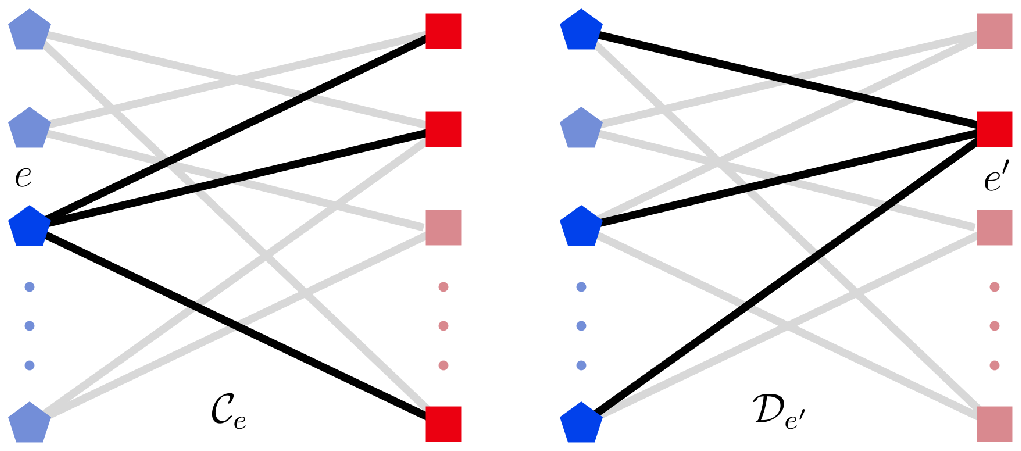}
  \normalfont
  \caption{\label{fig:space}(Color online) 
    Interpretation of the bipartite graph $G=\varphi_{\mathcal{L}^\prime}(\mathcal{L})$ 
    as given in Def.~\ref{def:map}. Each vertex $e$ identified with a \textit{deleted edge} 
    in $\mathcal{E}$ (\infig{-0.4}{0}{0.035}{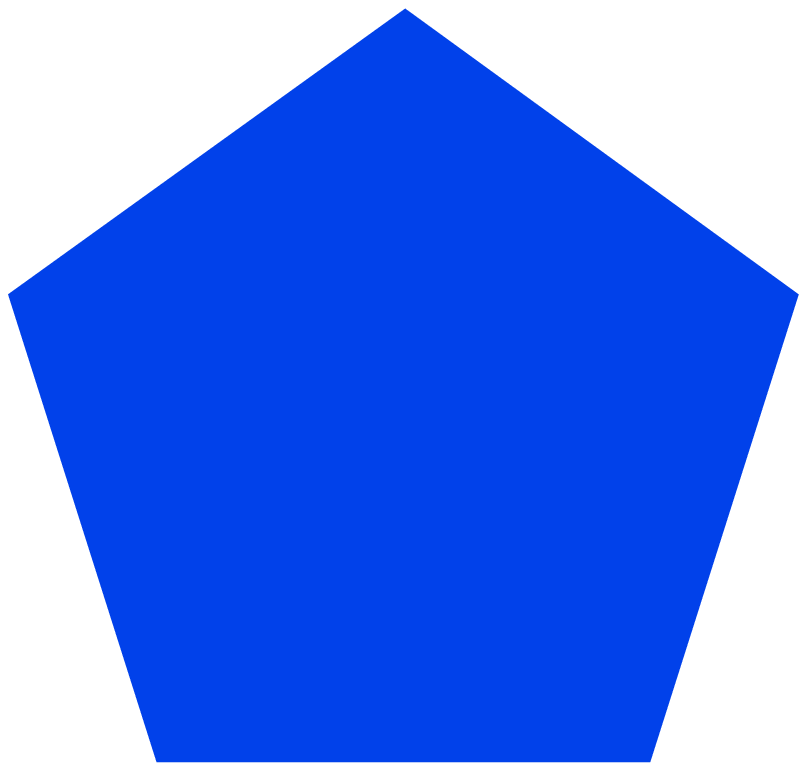}) encodes a \textit{fundamental cycle} $\mathcal{C}_e$ on $\mathcal{L}$ 
    with respect to $\mathcal{L}^\prime$ via its neighbours in $G$, viz. 
    $\mathcal{C}_e=N_e\cup\{e\}$. Conversely, each vertex $e^\prime$ identified with a 
    \textit{spanning tree edge} (\infig{-0.36}{0}{0.035}{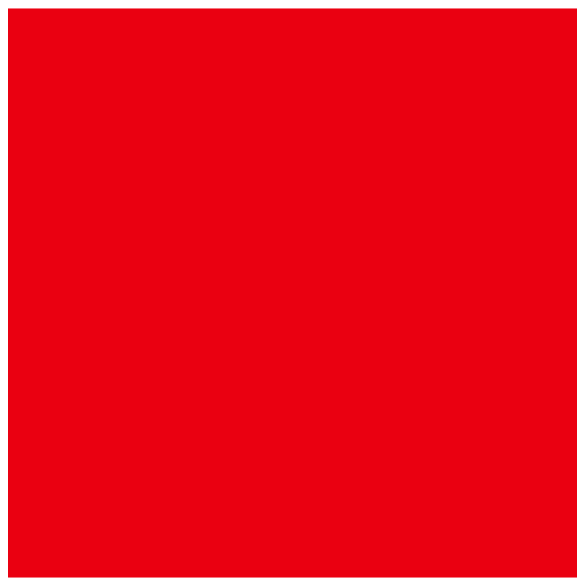}) encodes a \textit{fundamental cut} $\mathcal{D}_{e^\prime}$ 
    on $\mathcal{L}$ with respect to $\mathcal{L}^\prime$
    via its neighbours in $G$, viz. $\mathcal{D}_{e^\prime}=N_{e^\prime}\cup\{e^\prime\}$.
    This is a consequence of the duality between fundamental cuts and fundamental cycles.
  }
\end{figure}

Within this graph theoretical framework, the structure of 
$\varphi_{\mathcal{L}^\prime}(\mathcal{L})$ reads as follows.
First, we notice that $\varphi_{\mathcal{L}^\prime}(\mathcal{L})$ is bipartite, that is, 
there is a bipartition of its vertices, so that edges interconnect vertices
of different classes only. This bipartition is given by the set of vertices which belonged
to the spanning tree $\mathcal{L}^\prime$ as edges of $\mathcal{L}$, and its complement.
Therefore we may draw $\varphi_{\mathcal{L}^\prime}(\mathcal{L})$ 
schematically as depicted in Fig.~\ref{fig:space}. It is now straightforward to see that the 
neighbourhood $N_{e}$ ($N_{e^\prime}$) of vertices
$e\in\mathbb{E}\left(\mathcal{L}\right)\setminus \mathbb{E}\left(\mathcal{L}^\prime\right)$ 
($e^\prime\in\mathbb{E}\left(\mathcal{L}^\prime\right)$)
on $\varphi_{\mathcal{L}^\prime}(\mathcal{L})$ joined with $e$ ($e^\prime$) equals the 
fundamental cycle $\mathcal{C}_{e}$ (cut $\mathcal{D}_{e^\prime}$) on $\mathcal{L}$
with respect to $\mathcal{L}^\prime$.
We conclude that $\varphi_{\mathcal{L}^\prime}(\mathcal{L})$ \textit{encodes} the whole 
cycle space $\mathfrak{C}\left(\mathcal{L}\right)$,
and the cut space $\mathfrak{D}\left(\mathcal{L}\right)$ as well. 
In Fig.~\ref{fig:space} the highlighted subgraphs represent a fundamental cycle $\mathcal{C}_e$
and a fundamental cut $\mathcal{D}_{e^\prime}$ indexed by $e$ and $e^\prime$, respectively.
That both, $\mathfrak{C}\left(\mathcal{L}\right)$ and $\mathfrak{D}\left(\mathcal{L}\right)$, 
are encoded in the \textit{same} graph is not a conincidence but rather a necessity
since it is well known that there is a duality between fundamental cycles and fundamental 
cuts \cite{Diestel201010}; $\varphi_{\mathcal{L}^\prime}(\mathcal{L})$ is a geometrical
illustration of this duality. Since entanglement within the framework of graph states is 
characterised by edges between the qubits/vertices, the LC-equivalence of 
$\Ket{\varphi_{\mathcal{L}^\prime}(\mathcal{L})}$ to a given surface code ground state 
on $\mathcal{L}$ illustrates that the entanglement pattern of a surface code is 
basically determined by its cycle and cut space structure.

\subsection{Proof of Theorem 1}
\label{subsec:proof}

In this subsection the proof of Theorem \ref{thm:1} is presented in detail.
Readers primarily interested in the minimal non-local toric code settings may skip
this rather technical part and proceed with Section \ref{sec:minimal}.

\begin{figure*}[tbp]
  \tiny
  \textbf{A}\includegraphics[width=0.18\linewidth]{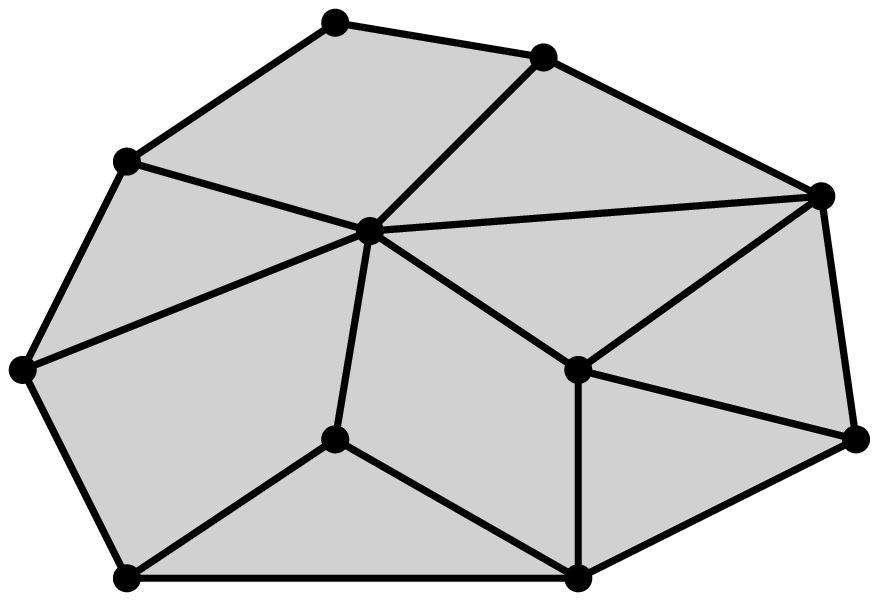}
  \textbf{B}\includegraphics[width=0.18\linewidth]{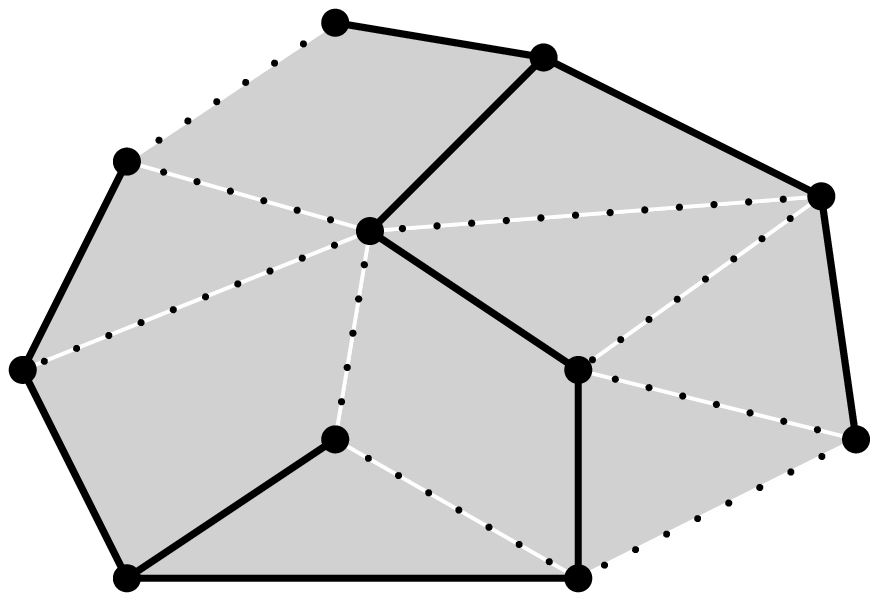}
  \textbf{C}\includegraphics[width=0.18\linewidth]{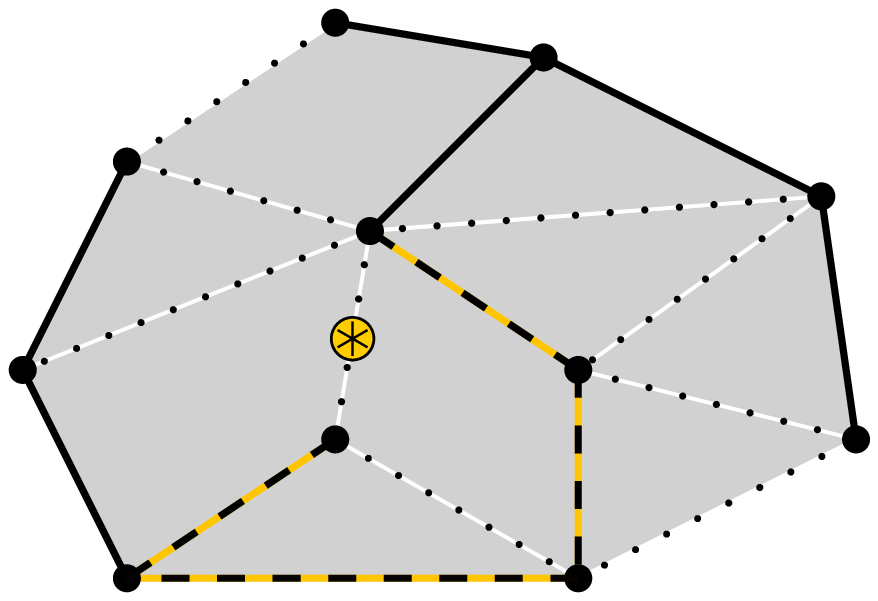}
  \textbf{D}\includegraphics[width=0.18\linewidth]{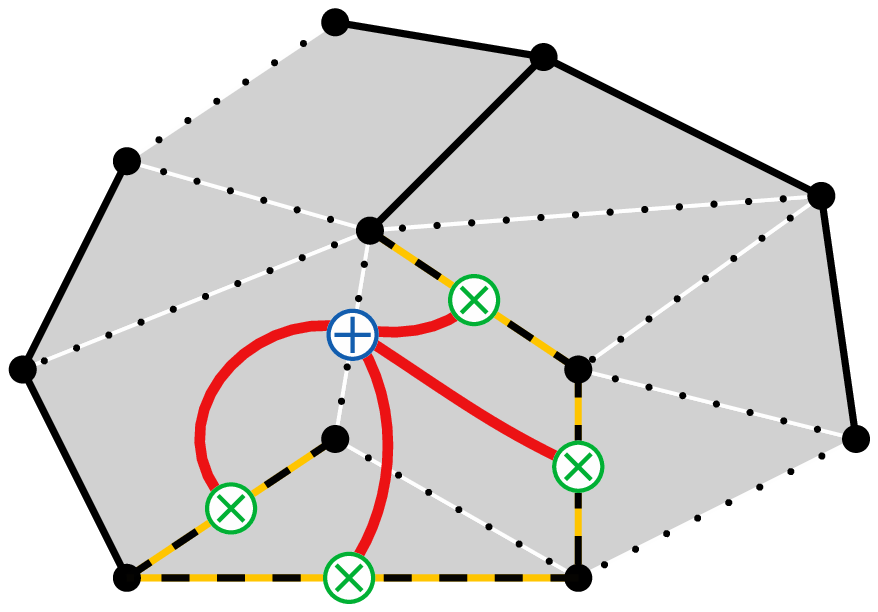}
  \textbf{E}\includegraphics[width=0.18\linewidth]{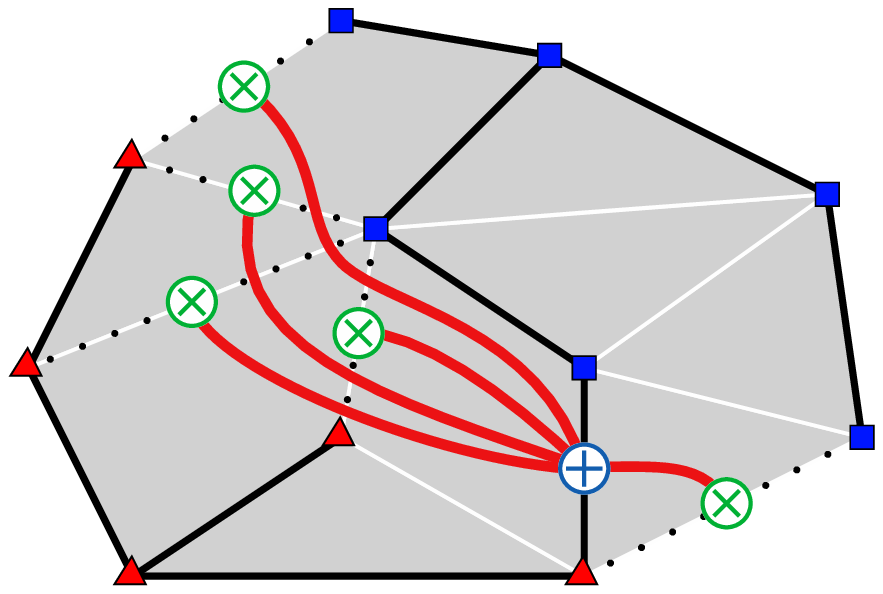}
  \caption{(Color online) 
    Construction of $G=\varphi_{\mathcal{L}^\prime}(\mathcal{L})$ as given in 
    Def.~\ref{def:map} and used in the proof of Theorem~\ref{thm:1}.
    Given a surface code (\textbf{A}) with arbitrary spanning tree (\textbf{B}). 
    Consider one of the (deleted) edges $e=\{p,q\}$ on which a Hadamard rotation was performed
    (denoted by \infig{-0.6}{-0.6}{0.08}{hadamard})
    and its unique path $\mathcal{C}_{pq}$ (denoted by 
    \infig{-0.0}{-0.8}{0.35}{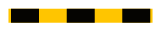}) on the spanning tree (\textbf{C}).
    The former $z$-type loop operator along $\mathcal{C}_{pq}+e$ acts as $\sigma^x$ 
    on the qubit \infig{0.0}{-0.2}{0.35}{xoperator} (due to the Hadamard rotation)
    and as $\sigma^z$ on the qubits denoted by \infig{0.0}{-0.2}{0.35}{zoperator}.
    This is a graph state stabilizer of the graph drawn with 
    \infig{0.0}{-0.4}{0.5}{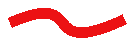} (\textbf{D}).
    On the other hand, consider one of the eges $e^\prime$ on the spanning tree (\textbf{E}).
    The induced bipartition $\mathbb{V}_r^{e^\prime}\,\dot\cup\,\mathbb{V}_b^{e^\prime}$ of 
    the vertices is denoted by \infig{0.2}{-0.4}{0.6}{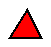}
    and \infig{0.2}{-0.2}{0.6}{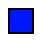}. The product of all (former) star operators 
    labeled by one of these sets acts nontrivially
    as $\sigma^x$ on the qubit \infig{0.0}{-0.2}{0.35}{xoperator} and as $\sigma^z$ on 
    the qubits denoted by \infig{0.0}{-0.2}{0.35}{zoperator}
    (due to the Hadamard rotations). This is a graph state stabilizer of the graph drawn 
    with \infig{0.0}{-0.4}{0.5}{wave_line} in (\textbf{E}).
  }
  \label{fig:proof}
  \normalsize
\end{figure*}

\begin{proof}[Proof of Theorem 1]
  
  First note that the basis states $\Ket{\vec v}$ of the protected space 
  are LC-equivalent since the topological quantum numbers $v_i$ are controlled by
  loop operators $X_k\in C_N^l$.
  Therefore, from the viewpoint of LU/LC-classification, w.l.o.g. assume $\vec v=(1,\dots,1)$.
  It follows that \textit{for each cycle} in $\mathcal{L}$ there is a $z$-type loop operator
  in the stabilizer $\mathcal{S}[\mathfrak{X}]$ (Fig.~\ref{fig:proof}~(A)).
  Let $\mathcal{L}^\prime$ be an arbitrary spanning tree and $\mathcal{E}$
  the set of edges such that $\mathcal{L}^\prime=\mathcal{L}-\mathcal{E}$ 
  (Fig.~\ref{fig:proof}~(B)). Apply \textit{Hadamard transformations} $H_e\in C_N^l$ 
  to the qubits identified with edges $e\in\mathcal{E}$.
  Formally this transformation reads $\Ket{\vec v}\mapsto H^{\mathcal{E}}\Ket{\vec v}$
  where $H^{\mathcal{E}}=\otimes_{e\in\mathcal{E}}H_e$.
  We now argue that $H^{\mathcal{E}}\Ket{\vec v}=\Ket{\varphi_{\mathcal{L}^\prime}(\mathcal{L})}$
  by showing that the new stabilizer
  $\mathcal{S}^\prime:=H^{\mathcal{E}}\mathcal{S}[\mathfrak{X}]\left(H^{\mathcal{E}}\right)^\dagger$
  is the graph state stabilizer described by $\varphi_{\mathcal{L}^\prime}(\mathcal{L})$.

  In the stabilizer picture, a Hadamard transformation is represented by the substitution
  $\sigma^x\,\leftrightarrow\,\sigma^z$.
  Consider one of these transformed qubits (\infig{-0.4}{-0.2}{0.08}{hadamard} in 
  Fig.~\ref{fig:proof}~(C)) and its edge $e=\{p,q\}\in \mathbb{E}(\mathcal{L})$
  ($p,q\in \mathbb{V}(\mathcal{L})$).
  The adjacent vertices $p$ and $q$ belong to both $\mathcal{L}$ and $\mathcal{L}^{\prime}$.
  Since $\mathcal{L}^\prime$ is a tree,
  there exists \textit{one and only one} path $\mathcal{C}_{pq}$ on 
  $\mathcal{L}^\prime$ connecting $p$ and $q$
  (\infig{0.0}{0.0}{0.35}{linedashed} in Fig.~\ref{fig:proof}~(C)). 
  Let $\sigma^z_e\otimes Z\left[\mathcal{C}_{pq}\right]$ be a combination of plaquette 
  operators in $\mathcal{S}[\mathfrak{X}]$ (note that $e+\mathcal{C}_{pq}$ is a 
  cycle in $\mathcal{L}$). This stabilizer operator is transformed under the Hadamard 
  transformation as follows
  \begin{eqnarray}
    K_{G}^{(e)}&=&H_e\sigma^z_e\otimes Z\left[\mathcal{C}_{pq}\right]H_e^\dagger
    =\sigma_e^x\otimes\bigotimes\limits_{i\in\mathcal{C}_{pq}}\sigma^z_i\,.
  \end{eqnarray}
  Obviously this is a graph state generator of 
  $\mathcal{S}\left[\varphi_{\mathcal{L}^\prime}(\mathcal{L})\right]$
  since in $\varphi_{\mathcal{L}^\prime}(\mathcal{L})$
  vertex $e$ is connected to all edges in $\mathcal{L}$ that belong to $\mathcal{C}_{pq}$
  (Fig.~\ref{fig:proof}~(D)).
  This argument holds for any edge in $\mathcal{E}$. 
  Therefore we find that the transformed stabilizer $\mathcal{S}^\prime$
  contains at least the graph state generators of 
  $\varphi_{\mathcal{L}^\prime}(\mathcal{L})$ for vertices in $\mathcal{E}$.
  It remains to show that the graph state generators
  for vertices $e^\prime\in \mathbb{E}\left(\mathcal{L}^\prime\right)$ can be 
  found in $\mathcal{S}^\prime$ as well.

  To this end consider an arbitrary qubit $e^\prime\in \mathbb{E}\left(\mathcal{L}^\prime\right)$
  in the spanning tree (\infig{-0.2}{0.0}{0.35}{xoperator} in Fig.~\ref{fig:proof}~(E)).
  The corresponding edge of the spanning tree cuts the latter into two parts.
  This construction yields a \textit{bipartition} of the vertex set
  $\mathbb{V}\left(\mathcal{L}^\prime\right)=\mathbb{V}_r^{e^\prime}\,\dot\cup\,\mathbb{V}_b^{e^\prime}$
  (red \infig{0.2}{0.0}{0.6}{small_triangle} and blue \infig{0.2}{0.0}{0.6}{small_square} 
  in Fig.~\ref{fig:proof}~(E)). Thus there are two types of edges in $\mathcal{E}$: 
  The first type connects vertices of the same partition and the second type interconnects 
  both sets $\mathbb{V}_b^{e^\prime}$ and $\mathbb{V}_r^{e^\prime}$.
  By definition of $\varphi_{\mathcal{L}^\prime}(\mathcal{L})$
  a vertex in $\mathcal{E}$ is connected to $e^\prime$ iff it is a type-two vertex.
  
  We claim that the following operator in $\mathcal{S}^\prime$
  is the graph state generator for $e^\prime$ 
  \begin{equation}
    K_{G}^{\left(e^\prime\right)}=\prod\limits_{s\in \mathbb{V}_r^{e^\prime}}A_s^\prime
    \quad\text{where}\quad
    A_s^\prime=H^{\mathcal{E}}A_s\left(H^{\mathcal{E}}\right)^\dagger\,.
  \end{equation}
  To see this, note that $\prod_{s\in \mathbb{V}_r^{e^\prime}}A_s^\prime$
  may affect a qubit $f$ in three different ways:   
  
  First, if $f$ belongs to an edge
  with adjacent vertices of \textit{the same} partition, nothing happens 
  since $\sigma^z\sigma^z=\mathds{1}=\sigma^x\sigma^x$.
  The second case occurs if the adjacent vertices belong to
  \textit{different} subsets \textit{and} $f\neq e^\prime$ 
  (\infig{-0.2}{0.0}{0.35}{zoperator} in Fig.~\ref{fig:proof}~(E)). 
  Since this case can only apply if $f\in\mathcal{E}$, a $\sigma^z$ operation is 
  performed on this qubit. The last case resembles the previous one, except for the 
  fact that $f=e^\prime$. Thus $\prod_{s\in \mathbb{V}_r^{e^\prime}}A_s^\prime$ 
  acts as $\sigma^x$ on $f$.

  To sum up: we found that $K_{G}^{\left(e^\prime\right)}$ acts as $\sigma^x$ on $e^\prime$, 
  as $\sigma^z$ on all vertices $f$ such that 
  $\left\{e^\prime,f\right\}\in \mathbb{E}(\varphi_{\mathcal{L}^\prime}(\mathcal{L}))$,
  and trivially otherwise. Therefore we showed that
  $\mathcal{S}\left[\varphi_{\mathcal{L}^\prime}(\mathcal{L})\right]\leq\mathcal{S}^\prime$.
  To show $\mathcal{S}\left[\varphi_{\mathcal{L}^\prime}(\mathcal{L})\right]=\mathcal{S}^\prime$ 
  we note that
  $\operatorname{rank}\mathcal{S}\left[\varphi_{\mathcal{L}^\prime}(\mathcal{L})\right]=
  N=\operatorname{rank}\mathcal{S}\left[\mathfrak{X}\right]=\operatorname{rank}\mathcal{S}^\prime$.

\end{proof}

\section{Minimal non-local toric code instances}
\label{sec:minimal}

\paragraph*{Setting the scene.}

In this section, we present minimal instances for the toric code on square
and triangular lattices which are \textit{not} LC-equivalent to any \textit{local} graph state.
We already provided a brief explanation and motivation for the notion of ``local'' in the 
introductory paragraphs. At this point, we shall give a more formal definition and 
relate the idea of locality to the toric code hamiltonian.

The latter is usually written as
\begin{equation}
  \label{eq:hamiltonian}
  H_\text{tcm}=-J_A\sum\limits_{s\in\mathbb{V}(\mathcal{L})}A_s-J_B
  \sum\limits_{p\in\mathbb{P}(\mathcal{L})}B_p
\end{equation}
where $J_A$ and $J_B$ define the excitation energies for electric and magnetic charges, 
respectively. Obviously the aforementioned protected space $\mathcal{PS}$ is also 
the ground state space of the above hamiltonian and the orthogonal states $\Ket{\vec v}$ 
constitute a basis of the latter. In the preliminaries the toric code was described
as a \textit{quantum code} within the stabilizer framework. However, considering 
the hamiltonian (\ref{eq:hamiltonian}), the toric code gains significance as an 
actual \textit{physical} system that may be implemented directly or by means of a 
quantum simulator. Crucial for the physical relevance of $H_\text{tcm}$ is its quasi-locality, 
that is, only quasi-local interactions -- represented by star and plaquette operators -- are 
required. Therefore, from a \textit{physical} point of view, the star and plaquette operators
are distiguished by their \textit{quasi-locality}, whereas from a \textit{mathematical} point of 
view all generating sets of $\mathcal{S}$ are equivalent.

Consequently, the set of star and plaquette operators gives rise to a physically motiviated 
adjacency relation on the set of qubits -- namely, two qubits $p$ and $q$ are \textit{vicinal} 
if both belong to the support of at least one star or plaquette operator, write 
$p\wr_\text{tc} q$. On the other hand, the relevance of graph states in experimental physics
is due to their scalability which in turn relies on their successive implementation by 
means of 2-qubit operations (see Eq.~(\ref{eq:graphstate2})). Therefore it seems appropriate 
to call two qubits \textit{vicinal in a graph state} if they are connected by an edge, 
write $p\wr_\text{gs} q$. The introduced symmetric relations $\wr$ encode physically motivated 
adjacencies \textit{with respect} to a particular class of multipartite entangled states.

After introducing the notion of vicinity, we may now ask the following question: Given the 
ground state $\Ket{\vec v}$ of a toric code, is there an LU-equivalent graph state 
$\Ket{G_\text{loc}}$, such that for all pairs of qubits 
$p\wr_\text{gs} q\,\Rightarrow\,p\wr_\text{tc} q$ holds. Provided such a graph state exists, 
we may apply those implementation techniques for graph states which gave rise to $\wr_\text{gs}$
(within the setup defined by the toric code) in order to generate its ground states. 
Conversely, if there is no such state, the considered toric code setup yields ground states 
which are fundamentally different from its equivalent graph states, that is, the toric 
code ground states cannot be implemented by means of graph states
without breaking the spatial structure of the toric code setup. At this point, the 
quasi-local property of $H_\text{tcm}$ becomes relevant. Thus we call toric code setups 
that exhibit at least one graph state $\Ket{G_\text{loc}}$ \textit{local} and $G_\text{loc}$ 
($\Ket{G_\text{loc}}$) itself a \textit{local graph (state)} with respect to the considered 
toric code setup. Hence a toric code is local if one can draw an LU-equivalent graph 
state \textit{without} ``long-range'' edges, that is, connecting qubits which cannot interact 
via star or plaquette operators in the toric code regime. E.g. see Fig.~\ref{fig:mintcm} 
for \textit{non-local} and Fig.~\ref{fig:example} for \textit{local} instances of toric 
code systems.


Small toric code systems tend to be local, e.g. the single toric code plaquette 
in Fig.~\ref{fig:intro}~(A) is represented by a local star graph. However, this property 
is likely to break down for more complex setups.
This motivates our search for the smallest \textit{non-local} toric code systems.
In doing so, we restrained the analysis to small systems on square and triangular lattices,
since experimental implementations are likely to use such tesselations.

\subsection{Methods}

\begin{figure}[tbp]
  \includegraphics[scale=0.12]{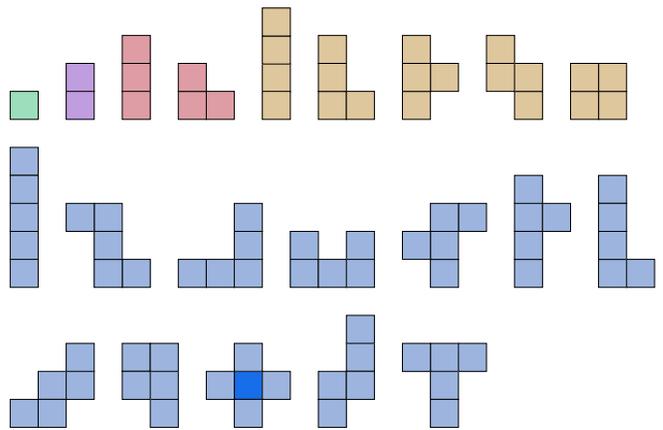}  
  \caption{\label{fig:polyomino}(Color online) 
  The free polyominos for $n=1,2,3,4,5$. Faces with at least one peripheral edge 
  are light-coloured. The first polyomino featuring a periphery-free cell (dark blue) is
  the $+$-shaped pentomino. It is the smallest configuration on the square lattice,
  such that the geometical construction yields only non-local graph states.}
\end{figure}

Our approach is as follows. Due to the restriction to a particular tesselation
(let us consider the square lattice exemplarily) we are looking for connected subsets of 
squares on the square lattice. To this end we may consider only setups with 
\textit{edge-connected} squares since \textit{vertex-connected} structures yield separable
ground states or (if there is a loop through the vertices) degenerate ground states; 
the latter requiring a reduction to a non-degenerate setup by
``filling'' the loop, anyway \footnote{This is equivalent to the choice 
$\Ket{\vec v}=\Ket{1,1,\dots,1}$.}. Such structures are known in combinatorics as 
\textit{polyominos} (on the square lattice) and \textit{polyiamonds} (on the triangular lattice).
The polyominos for up to $n=5$ squares are depicted in Fig.~\ref{fig:polyomino}.
Application of Theorem \ref{thm:1} yields local graph states for all depicted setups devoid 
of ``inner'' squares -- provided a suitable spanning tree is
used for the construction. Actually, by means of Theorem \ref{thm:1} this can be worked 
out mentally. Subsequently there is a smallest setup left, such that the geometric 
transformation yields only non-local graphs. For the square lattice this would be the 
$+$-shaped pentomino ($n=5$) with dark coloured inner square, as depicted in 
Fig.~\ref{fig:polyomino}.

However, the class of geometrically constructed graph states is generally
a proper subset of the whole LC-class containing \textit{all} LC-equivalent graph 
states \footnote{This can be seen easily since any non-trivial LC-class 
  $\left[G\right]_{\mathrm{LC}}$ contains graphs with
  $3$-cycles but $\varphi_{\mathcal{L}^\prime}(\mathcal{L})$ is a bipartite graph by 
  construction, and therefore features only cycles with an \textit{even} number of edges.}.
To fix this issue, we computed the whole LC-class and searched for local representatives.
To this end, an algorithm described in \cite{1_hein_EIGSAIA_S} and originally 
derived in \cite{Bouchet1991,Bouchet199375} was employed, according to which two 
graphs $G$ and $G^\prime$ with adjacency matrices \footnote{That is, matrices with 
  one column/row for each vertex, a $1$ at the corresponding position if two vertices 
  are connected and a $0$ otherwise.} $\mathbf{\Gamma}$ and $\mathbf{\Gamma}^\prime$
are equivalent under local complementations iff there exist diagonal matrices
$\mathbf{A},\mathbf{B},\mathbf{C},\mathbf{D}\in\mathbb{F}_2^{N\times N}$, 
such that the non-linear condition $\mathbf{A}\mathbf{D}+\mathbf{B}\mathbf{C}=\mathbf{E}$
and the system of linear equations
\begin{equation}
  \label{eqn:ABCDlin}
  \left(\mathbf{\Gamma}\mathbf{B}+\mathbf{D}\right)\mathbf{\Gamma}^\prime+
  \left(\mathbf{\Gamma}\mathbf{A}+\mathbf{C}\right)=\mathbf{0}
\end{equation}
are satisfied over the Galois field $\mathbb{F}_2$ ($\mathbf{E}$ denotes the identity matrix).
This condition was implemented using \texttt{Mathematica} and facilitated the generation 
of the whole LC class $[G]_{\mathrm{LC}}$ for a given graph $G$.
However, this algorithm is inefficient in the sense that
computation time increases exponentially with the number of vertices.
To our knowledge, no efficient (i.e. polynomial) algorithm has yet been found.
Unfortunately, the exponential scaling behaviour renders the LC-class generation 
for the $+$-shaped setup (i.e. $N=16$ vertices/qubits) unfeasible on a regular personal 
computer, whereas the triangular tetriamond depicted in Fig.~\ref{fig:mintcm}~(B) 
(i.e. $N=9$ vertices/qubits) requires only few minutes to generate the whole LC-class 
and seek for local graphs. To deal with this (rather technical) problem, we conceived 
a method to deduce the non-locality of certain toric code setups from smaller ones. 
The details concerning this method are technical and purely graph theoretic. 
We therefore omit them at this point and refer the interested reader to Appendix \ref{appendixI}.

\subsection{Results}
As described above, the smallest setup on the square lattice without \textit{geometrically} 
producable local graph state is the $+$-shaped pentomino ($N=16$), as depicted in 
Fig.~\ref{fig:mintcm}~(A). The aforementioned reduction method yields the smaller ($N=8$)
setup depicted in Fig.~\ref{fig:mintcm}~(C). We proved the non-locality of the latter 
by generating the whole LC-class and probing for local graphs. Since the result was 
negative, we conclude that the system depicted in Fig.~\ref{fig:mintcm}~(C) as well as
the $5$-plaquette setup in Fig.~\ref{fig:mintcm}~(A) are non-local. Therefore, on the 
square lattice, the $+$-shaped pentomino with $N=16$ qubits is the smallest setup 
without (LC-)equivalent \textit{local} graph state.

On the triangular lattice the triangle shaped tetriamond in Fig.~\ref{fig:mintcm}~(B) 
is the smallest one yielding only non-local, \textit{geometrically} constructed graph 
states. Furthermore, the generation of all LC-equivalent graphs yields only non-local ones.
Thus, on the triangular lattice, the triangle-shaped tetriamond with $N=9$ qubits is 
the smallest setup without (LC-)equivalent \textit{local} graph state.

\begin{acknowledgments}
We acknowledge support by the Deutsche Forschungsgemeinschaft (DFG) within SFB
/ TRR 21, National Science Foundation under Grant No. NSF PHY05-51164 and
the Aspen Center for Physics with NSF grant 1066293.
\end{acknowledgments}


\bibliographystyle{h-physrev}
\bibliography{bibtex/mybib}

\appendix

\section{Reduction of surface code systems}
\label{appendixI}
\small

\subsection{Structure of \texorpdfstring{$\varepsilon$}{Epsilon}-symmetric LC-classes}

In this section, some properties of LC-classes of simple graphs are presented that may be used
to decide whether certain surface code systems are local. The following statements are purely
graph theoretic.

\begin{Lemma}[$\varepsilon$-Symmetry]
  \label{lemma:sym}
  Let $[G]_{\mathrm{LC}}$ be the LC-class of a simple graph $G$ such that there 
  exist representatives with at least one leaf. If $a$ and $b$ denote the outer and 
  inner vertex of a leaf, then the whole LC-class is invariant under the exchange 
  of $a$ and $b$. We denote this symmetry by
  \begin{equation}
    \varepsilon_{ab}\,:\,\mathfrak{E}(V)\,\rightarrow\,\mathfrak{E}(V)
  \end{equation}
  and call $[G]_{\mathrm{LC}}$ \textbf{$\varepsilon$-symmetric} with respect to $a$ and $b$.
  Formally this reads $\varepsilon_{ab}[G]_{\mathrm{LC}}=[G]_{\mathrm{LC}}$.
\end{Lemma}
Here $\mathfrak{E}(V)$ denotes the \textit{edge space} over the vertex set $V$, that is,
the set of all simple graphs with vertices $V$ equipped with the binary addition of edges.
\begin{proof}
  Let $G_{ab}$ be a representative with a leaf and $a$ be the outer vertex. 
  Then it is easy to verify that the two local complementations
  \begin{equation*}
    G_{ba}=\tau_a\tau_b G_{ab}
  \end{equation*}
  yield a graph $G_{ba}$ such that $b$ is the outer vertex and all
  edges that connected $b$ in $G_{ab}$ with vertices $V\setminus \{a,b\}$ now emanate from $a$.
  In other words: The graph $G_{ba}$ is obtained from $G_{ab}$ by exchanging the 
  vertices $a$ and $b$. Hence the symmetry $\varepsilon_{ab}G_{ab}=G_{ba}$ is realized 
  by two local complementations. Now consider an arbitrary element 
  $H_{ab}\in [G]_{\mathrm{LC}}$. $H_{ab}$ is reachable from $G_{ab}$
  by subsequent application of local complementations. If $\tau(a,b)$ denotes this chain 
  of LC-operations, then let $\tau(b,a)$ be the chain obtained by the 
  substitution $\tau_a\leftrightarrow\tau_b$. Then
  $H_{ba}=\tau(b,a)G_{ba}=\tau(b,a)\varepsilon_{ab}G_{ab}=\tau(b,a)\tau_a\tau_bG_{ab}$
  differs from $H_{ab}$ by a permutation of $a$ and $b$;
  we write $H_{ba}=\varepsilon_{ab}H_{ab}\in [G]_{\mathrm{LC}}$. Thus we find
  $\varepsilon_{ab}^2[G]_{\mathrm{LC}}\subseteq\varepsilon_{ab}[G]_{\mathrm{LC}}\subseteq[G]_{\mathrm{LC}}$
  and since $\varepsilon_{ab}^2=\mathds{1}$ it
  follows $\varepsilon_{ab}[G]_{\mathrm{LC}}=[G]_{\mathrm{LC}}$.
\end{proof}

The symmetry originally derived for the special leaf-graphs $G_{ab}$ and $G_{ba}$ 
``spreads'' through the whole equivalence class and leads to some useful properties 
regarding LC-classes represented by such graphs. Before analysing this structure, let 
us prove the following statement.

\begin{Lemma}[Deletion of leafs]
  \label{lemma:delete}
  Let $G_{ab}\in\mathfrak{E}(V)$ be a simple graph with at least one
  leaf; $a$ denotes the outer and $b$ the inner vertex. Consider the subgroup
  \begin{equation}
    \mathcal{T}_a:=
    \operatorname{span}\left\{\tau_i\,:\,\mathfrak{E}(V)\,\rightarrow\,
      \mathfrak{E}(V)\,|\,i\in V\setminus \{a\}\right\}
  \end{equation}
  of all local complementations on vertices $V\setminus \{a\}$. Then it holds
  \begin{equation}
    \forall\,\tau\in\mathcal{T}_a\,:\,\tau G_{ab}-a=\tau \left(G_{ab}-a\right)\,.
  \end{equation}
\end{Lemma}

In words: Deleting the leaf ($a$ and its adjacent edge) from $G_{ab}$ and transforming 
this graph by a sequence of local complementations $\tau$ acting on vertices 
$V\setminus \{a\}$ yields the same graph as if these operations were performed 
on $G_{ab}$ (without deleting the leaf) and deleting $a$ with all its adjacent
edges (which may be more than one) subsequently.

\begin{proof}
  It is easy to see that all changes regarding edges between vertices in 
  $V\setminus\{a\}$ occur in the same way, whether there is an edge to $a$ 
  or not. Therefore the only differences between $\tau G_{ab}$ and
  $\tau \left(G_{ab}-a\right)$ affect edges from vertices in $V\setminus\{a\}$ 
  to $a$ itself. Since these edges are deleted ($\tau G_{ab}-a$), there is no 
  difference left and it follows $\tau G_{ab}-a=\tau \left(G_{ab}-a\right)$.
\end{proof}

Lemma \ref{lemma:delete} allows statements about the LC-class $[G_{ab}-a]_{\mathrm{LC}}$.
More precisely: All graphs in $[G_{ab}]_{\mathrm{LC}}$ that are
reachable from $G_{ab}$ by concatenations of local complementations in $\mathcal{T}_a$
yield graphs in $[G_{ab}-a]_{\mathrm{LC}}$ by deleting $a$ and its adjacent edges.
If one can show that all graphs in $[G_{ab}]_{\mathrm{LC}}$ may be linked to each other
by operations in such subgroups, this yields a procedure to compute
LC-classes of special subgraphs.

This idea leads to the following proposition for simple graphs with leafs:
\begin{Proposition}[$\varepsilon$-symmetric LC-classes]
  \label{prop:sub}
  Let $G_{ab}$ be a simple graph with at least one leaf where $a$ and $b$ denote the outer and
  inner vertex, respectively.

  Then for all $H\in [G_{ab}]_{\mathrm{LC}}$ there 
  exists $H^\prime_a\in [G_{ab}-a]_{\mathrm{LC}}$ or
  $H^\prime_b\in [G_{ba}-b]_{\mathrm{LC}}$ such that 
  $H-a=H^\prime_a$ or $H-b=H^\prime_b$, respectively.
\end{Proposition}

In other words: If an LC-class features a leaf-graph representative with outer 
and inner vertex $a$ and $b$, then deleting of one of the latter (in some cases it 
does not matter which one is deleted) yields a graph
in $[G_{ab}-a]_{\mathrm{LC}}$ or $[G_{ba}-b]_{\mathrm{LC}}$. 
Consequently, any graph in $[G_{ab}]_{\mathrm{LC}}=[G_{ba}]_{\mathrm{LC}}$ features at least one subgraph in
$[G_{ab}-a]_{\mathrm{LC}}$ or $[G_{ba}-b]_{\mathrm{LC}}$.

\begin{proof}

  Let $G_{ab}$ be a representative with outer and inner vertex $a$ and $b$. 
  Then one can partition the LC-class $[G_{ab}]_{\mathrm{LC}}$ in the following way:
  \begin{itemize}
  \item All graphs with a leaf belong to $A$ iff $a$ is the outer and $b$ the inner vertex.
    Therefore $G_{ab}\in A$.
  \item All graphs with a leaf belong to $B$ iff $b$ is the outer and $a$ the inner vertex.
    Therefore $G_{ba}\in B$.
  \item All graphs $H$ where $\{a,b\}\in \mathbb{E}(H)$ and for all $v\in \mathbb{V}(H)\setminus\{a,b\}$
    it holds
    \begin{equation*}
      \{a,v\}\in \mathbb{E}(H)\,\Leftrightarrow\, \{b,v\}\in \mathbb{E}(H)
    \end{equation*}
    belong to $C$.
    Due to the second statement, these graphs are totally symmetric with respect to $a$ and $b$,
    $\varepsilon_{ab}H=H$, that is.
  \item All graphs $H$ where $\{a,b\}\notin \mathbb{E}(H)$ and for all $v\in \mathbb{V}(H)\setminus\{a,b\}$ 
    it holds
    \begin{equation*}
      \{a,v\}\in \mathbb{E}(H)\,\Leftrightarrow\, \{b,v\}\in \mathbb{E}(H)
    \end{equation*}
    belong to $D$.
    These graphs are totally symmetric; the only difference to type-$C$ graphs
    is the missing edge between $a$ and $b$.
  \end{itemize}
  
  \begin{figure}[htbp]
    \centering
    \vspace{0pt}
    \includegraphics[width=1.0\linewidth]{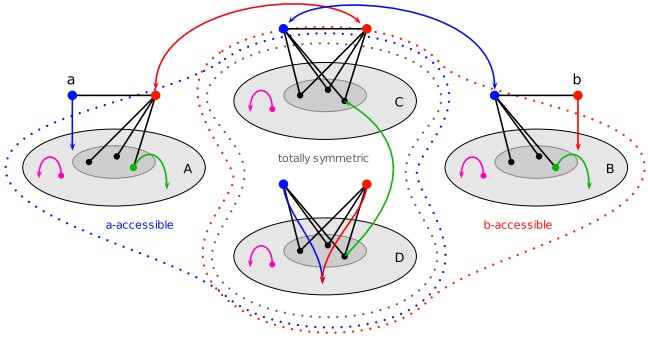}  
    \caption{\label{fig:lc_classes}(Color online) 
      Diagrammatic sketch showing the structure of $\varepsilon$-symmetric LC-classes as
      used in the proof of Proposition \ref{prop:sub}.
      Arrows denote different classes of transformations according to their color:
      $\gamma$-operations (magenta), $\xi$-operations (green), 
      $a$-operations (blue) and $b$-operations (red).
      A detailed description is given in the text.
    }
  \end{figure}

  These four types of graphs are depicted in Fig.~\ref{fig:lc_classes}. 
  Note that the number of edges
  between $a$ or $b$ and the vertices $\mathbb{V}(G)\setminus\{a,b\}$ (denoted by light grey discs)
  is \textit{not} constant within one class in general. That the described classification of graphs is a 
  \textit{partition}, i.e. $[G_{ab}]_{\mathrm{LC}}=A\,\dot\cup\,B\,\dot\cup\,C\dot\cup\,D$ 
  has to be shown. To this end, we introduce a classification of local complementations 
  $\tau_v\in\mathcal{T}$ which depends on the graph they operate on.
  First, we call $\tau_a$ and $\tau_b$ an \textit{$a$- and $b$-operation}, respectively.
  Secondly, local complementations $\tau_v$ (acting on $G$) where $v\in \mathbb{V}(G)\setminus\{a,b\}$ 
  and $\{a,v\},\{b,v\}\notin \mathbb{E}(G)$ are called \textit{$\gamma$-operations} and
  LC-operations $\tau_v$ (acting on $G$) where $v\in \mathbb{V}(G)\setminus\{a,b\}$ 
  and $\{a,v\}\in \mathbb{E}(G)$ or $\{b,v\}\in \mathbb{E}(G)$ are called \textit{$\xi$-operations}. 
  Obviously every local complementation $\tau_v$, $v\in \mathbb{V}(G)$, belongs 
  (depending on the graph it is applied to) to one of these four classes.
  Let us examine how these four classes of local complementations 
  and the four classes of graphs introduced above collude:

  \begin{itemize}
  \item First, note that $\gamma$-operations cannot alter the class a graph belongs to ($A$,$B$,$C$ or $D$) since
    all edges used for the classification are unaffected by $\gamma$-operations. In
    Fig.~\ref{fig:lc_classes} this is denoted by magenta arrows pointing from each set into itself.
  \item Clearly, $a$-operations act trivially on $A$. Applied to an element of $C$, 
    it deletes all edges between $b$ and $\mathbb{V}(G)\setminus\{a,b\}$ and yields a graph 
    that belongs to $B$. Since $\tau_a^2=\mathds{1}$, this is true for the opposite 
    direction as well. Finally, graphs that belong to $D$ cannot leave this set
    since no edge between $a$ and $b$ is created. These connections are denoted by blue
    arrows in Fig.~\ref{fig:lc_classes}.
  \item Due to the symmetry with respect to $a$ and $b$, all statements for $\tau_a$ hold for
    $\tau_b$ (i.e. $b$-operations) as well. 
    These relations are denoted by red arrows in Fig.~\ref{fig:lc_classes}. Note that $\tau_a$ equals
    $\tau_b$ if they are applied to elements in $D$.
  \item $\xi$-operations act on vertices denoted by dark grey circles in Fig.~\ref{fig:lc_classes}
    (vertices that are connected to $a$ or $b$ or both of them). Obviously such operations cannot
    manipulate the leaf. 
    Therefore $\xi$-operations cannot leave the sets $A$ and $B$ 
    (but they may alter the number of edges connecting $a$ or $b$ to vertices 
    in $\mathbb{V}(G)\setminus\{a,b\}$). Acting on elements in $C$, a
    $\xi$-operation deletes the edge between $a$ and $b$, hence converting type-$C$ graphs to
    type-$D$ graphs and vice versa. These relations are denoted by green arrows in Fig.~\ref{fig:lc_classes}.
  \end{itemize}

  Consider the following situation: Starting from $G_{ab}\in A$ one tries to generate as much
  elements in $[G_{ab}]_{\mathrm{LC}}$ as possible by using local complementations 
  in $\mathcal{T}_a$. Therefore one is allowed to use $b$- $\gamma$- and $\xi$-operations 
  to generate graphs. In terms of Fig.~\ref{fig:lc_classes} the ``path'' cannot use 
  the \textit{blue} arrows. Note that the only
  $a$-operation that cannot be omitted connects $C$ and $B$ (in $A$ it acts trivially and in
  $D$ it may be replaced by a $b$-operation). We call graphs in this orbit $\mathcal{T}_aG_{ab}$
  \textit{a-accessible}. In conclusion it 
  holds $\mathcal{T}_aG_{ab}\subseteq [G]_{\mathrm{LC}}\setminus B$.
  Analogue arguments yield $\mathcal{T}_bG_{ba}\subseteq [G]_{\mathrm{LC}}\setminus A$ 
  for \textit{b-accessible} graphs starting from $G_{ba}\in B$.

  We are now going to show that the more stricter relations 
  $\mathcal{T}_aG_{ab}= [G]_{\mathrm{LC}}\setminus B$
  and $\mathcal{T}_bG_{ba}= [G]_{\mathrm{LC}}\setminus A$ hold.
  To this end, consider an arbitrary ``path'' jumping from class to class, w.l.o.g. starting
  at $G_{ab}$ and ending at $H\in[G]_{\mathrm{LC}}\setminus B$, formally described by
  \begin{equation*}
    H=\prod\limits_{i}\tau_i G_{ab}\quad\text{where}\quad \tau_i\in\mathcal{T}
  \end{equation*}
  as depicted in Fig.~\ref{fig:lc_jump}~(A). 
  In general there will be some $a$-operations in this chain. Since neither $G_{ab}$ nor
  $H$ are in $B$ there is an even number of them (in Fig.~\ref{fig:lc_jump}~(A) there are two,
  denoted by blue jumps). We are restricted to non-$a$-operations in $\mathcal{T}_a$, though.
  We can adhere to this restriction by ``swapping'' the $a$-jumps and the 
  corresponding $\gamma$- and $\xi$-operations in $B$ to $b$-jumps and the 
  same $\gamma$- and $\xi$-operations in $A$. This is shown in Fig.~\ref{fig:lc_jump}~(B).
  
  \begin{figure}[htbp]
    \centering
    \includegraphics[width=1.0\linewidth]{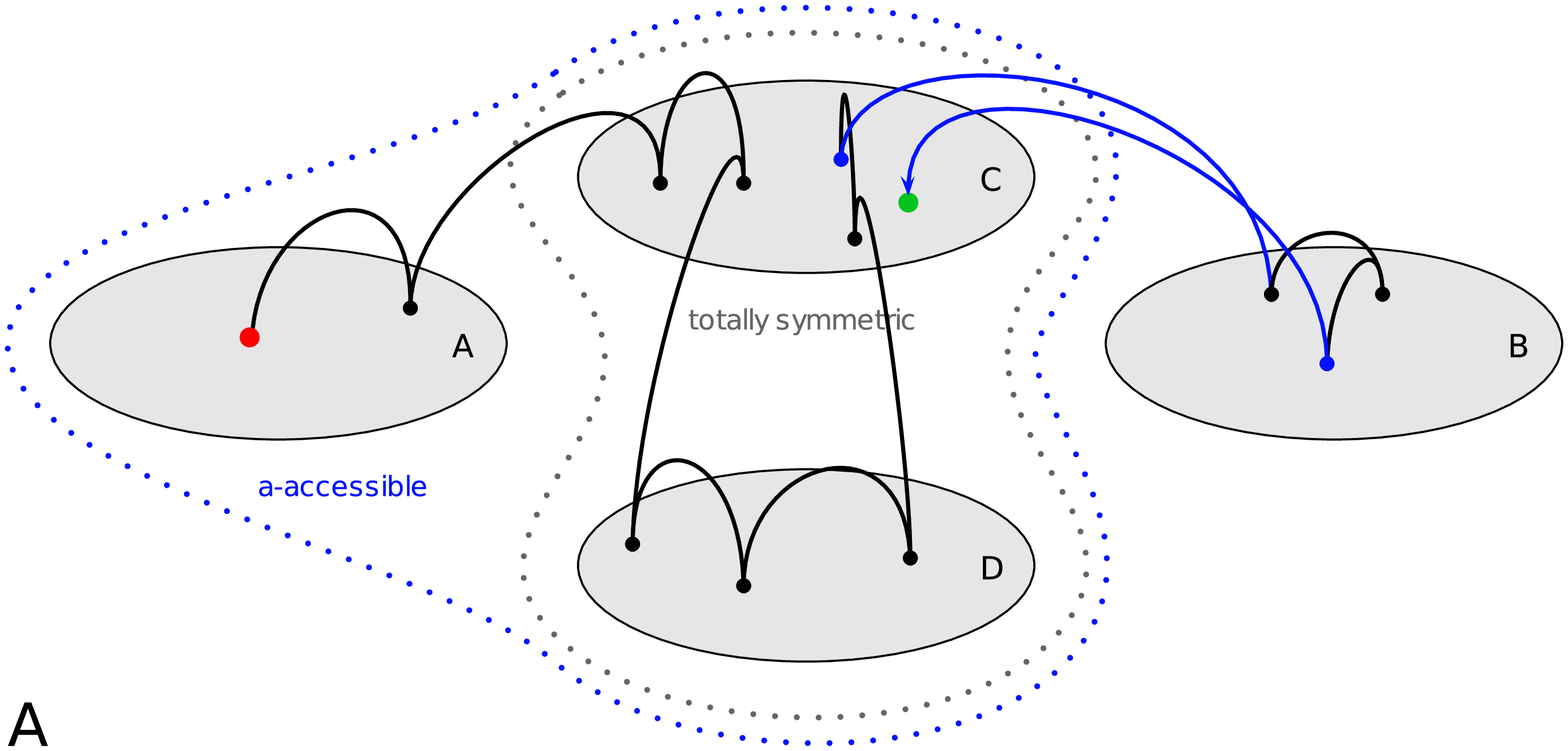}  
    \includegraphics[width=1.0\linewidth]{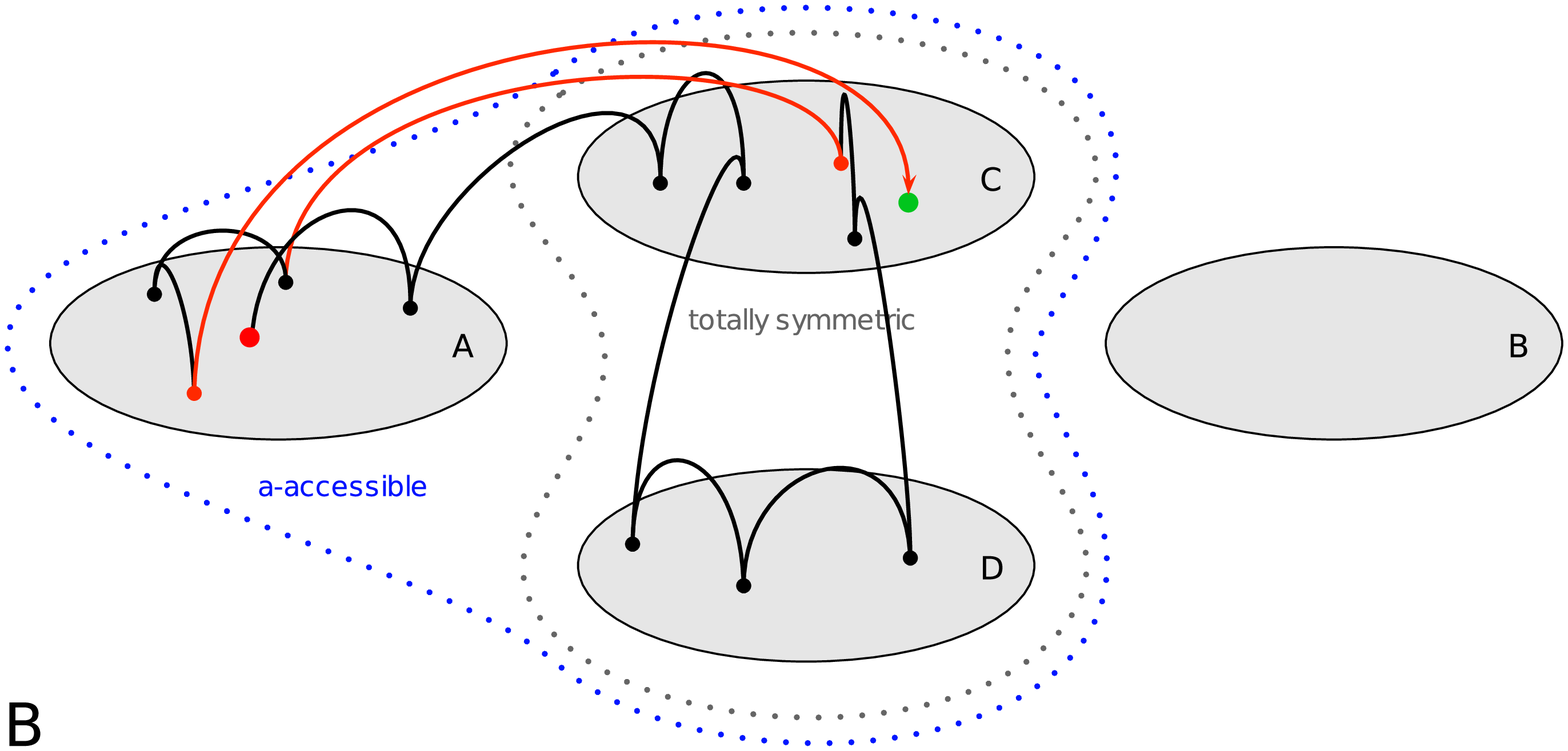}  
    \caption{\label{fig:lc_jump}(Color online) 
      An arbitrary path from $G_{ab}$ (red dot) to $H$ (green dot) is depicted in (\textbf{A}). 
      The ``forbidden'' (blue) paths using $a$-operations
      are swapped to $A$ (red jumps) using $b$-operations instead, see (\textbf{B}). Now the whole path 
      from $G_{ab}$ to $H$ is $a$-accessible. A detailed description is given in the text.
    }
  \end{figure}

  It is easy to see that jumping to $A$ by $b$-operations instead and subsequently applying 
  \textit{the same} $\gamma$- and $\xi$-operations yields a graph in $A$ that resembles the graph obtained
  in $B$ before jumping back to $C$ with the \textit{only} difference that $a$ and $b$ are permuted.
  However, since the second application of $\tau_b$ yields a totally symmetric graph 
  in $C$, this graph has to be the same one as formerly obtained by the second $\tau_a$ operation.

  Consequently we found a procedure to reach any graph 
  $H\in[G]_{\mathrm{LC}}\setminus B$ by local complementations in $\mathcal{T}_a$ 
  from $G_{ab}$ and, due to symmetry, each $H^\prime\in[G]_{\mathrm{LC}}\setminus A$
  by exclusive application of $\mathcal{T}_b$ operations from 
  $G_{ba}$ \footnote{It is noteworthy that elements in $A$ may be reached \textit{only} 
    from $G_{ab}$ and elements in $B$ \textit{only} from $G_{ba}$, whereas this procedure 
    shows that elements in $C$ and $D$ may be reached from both $G_{ab}$ and $G_{ba}$ 
    using $\mathcal{T}_a$ or $\mathcal{T}_b$, respectively.}.  
  Thus we showed 
  $\mathcal{T}_bG_{ba}= [G]_{\mathrm{LC}}\setminus A$ and
  $\mathcal{T}_aG_{ab}= [G]_{\mathrm{LC}}\setminus B$.

  To conclude the proof, recall Lemma \ref{lemma:delete}. Consider $H\in [G]_{\mathrm{LC}}$. 
  Assume without loss of generality that $H\notin B$ (the following argument holds 
  for $H\notin A$ analogously). Then we showed that $H\in \mathcal{T}_aG_{ab}$ 
  where $G_{ab}$ is a leaf-graph with outer vertex $a$. Choose $\tau\in\mathcal{T}_a$ 
  such that $H=\tau G_{ab}$. According to Lemma \ref{lemma:delete} it follows that
  $H-a=\tau G_{ab}-a=\tau\left(G_{ab}-a\right)$. Set 
  $H_a^\prime:=\tau\left(G_{ab}-a\right)\in [G_{ab}-a]_{\mathrm{LC}}$
  and it follows $H-a=H_a^\prime\in [G_{ab}-a]_{\mathrm{LC}}$. Which was to be proven.

\end{proof}

\subsection{Application to the locality problem}

In order to apply the graph theoretic knowledge about $\varepsilon$-symmetric LC-classes 
introduced above, we have to define in graph theoretic terms the notion of  
``local surface codes'' and ``local graph states''.

\begin{Definition}[Adjacency relation, Local graph]
  \label{def:locality}
  Let $V$ be an arbitrary finite set.
  An \textbf{adjacency relation}
  $\Lambda$ on $V$ is an \textbf{irreflexive} and \textbf{symmetric}
  relation, i.e. there is no $v\in V$ such that $v\sim_\Lambda v$ and 
  for each pair $v,w\in V$ it holds
  $v\sim_\Lambda w\,\Rightarrow\, w\sim_\Lambda v$.
  An adjacency relation is represented by an \textbf{adjacency matrix} $\mathbf{\Lambda}$
  where $\Lambda_{v,w}=1$ if $v\sim_\Lambda w$ and $\Lambda_{v,w}=0$ otherwise.
  Therefore $\Lambda$ defines a simple graph $G_{\Lambda}$ 
  with adjacency matrix $\mathbf{\Lambda}$.
  A simple graph $G_{\mathrm{loc}}$ is called \textbf{local} with respect to $\Lambda$ if
  it is a subgraph of $G_{\Lambda}$.
\end{Definition}

Given a surface code $\mathcal{L}$ (more precisely: a 2-cell embedding
$\mathfrak{X}_{\mathcal{L},\Sigma}$ of $\mathcal{L}$ in $\Sigma$),
the stabilizer generators $A_s$ and $B_p$ 
induce an adjacency relation $\Lambda$ on the set of edges (or qubits) 
$\mathbb{E}(\mathcal{L})$ as follows. Two qubits $e,f\in \mathbb{E}(\mathcal{L})$ are adjacent, 
$e\sim_\Lambda f$, if and only if there is a common face $p\in \mathbb{P}(\mathcal{L})$ or 
a common vertex $s\in \mathbb{V}(\mathcal{L})$ such that $e,f\in p$ or $e,f\in s$, respectively.
Note that by this definition of $\Lambda$, \textit{local graphs} $G_{\mathrm{loc}}$ in
the sense of Def. \ref{def:locality} are the very graphs describing 
\textit{local graph states} $\Ket{G_{\mathrm{loc}}}$ according to the definition given in
Section \ref{sec:minimal}.

Since we want to reduce surface code setups to smaller ones by removing edges/qubits,
a method to compare two setups with respect to their adjacency relations is needed.
Therefore we introduce the notion of \textit{strictness}.

\begin{Definition}[Strictness]
  \label{def:strict}
  Let $\mathcal{L}_1$ and $\mathcal{L}_2$ be two surface codes with edges (qubits)
  $\mathbb{E}\left(\mathcal{L}_2\right)\subseteq \mathbb{E}\left(\mathcal{L}_1\right)$.
  Furthermore denote by $\Lambda_1$ and $\Lambda_2$ the adjacency relations as defined above.
  They define simple graphs $G_{\Lambda_1}$ and $G_{\Lambda_2}$ with vertices
  $\mathbb{E}\left(\mathcal{L}_1\right)$ and $\mathbb{E}\left(\mathcal{L}_2\right)$, respectively.
  $\Lambda_1$ is called \textbf{stricter} than $\Lambda_2$ 
  (write $\Lambda_1 \succcurlyeq \Lambda_2$) if 
  $G_{\Lambda_1}\left[\mathbb{E}(\mathcal{L}_2)\right]\subseteq G_{\Lambda_2}$.
\end{Definition}

Here $G_{\Lambda_1}\left[\mathbb{E}(\mathcal{L}_2)\right]$ denotes the subgraph
of $G_{\Lambda_1}$ on the vertices
$\mathbb{E}(\mathcal{L}_2)\subseteq \mathbb{E}\left(\mathcal{L}_1\right)$.
Definition \ref{def:strict} and Proposition \ref{prop:sub} 
enable us to state and prove the following theorem, which is the main result
of this section:

\begin{Theorem}[Surface code reduction]
  \label{theo:reduction}
  Let $\mathcal{L}$, $\mathcal{L}_a$ and $\mathcal{L}_b$ be surface code systems
  with the property that there exist qubits $a,b\in \mathbb{E}(\mathcal{L})$ such that
  $\mathbb{E}(\mathcal{L}_a)=\mathbb{E}(\mathcal{L})\setminus\{a\}$ and 
  $\mathbb{E}(\mathcal{L}_b)=\mathbb{E}(\mathcal{L})\setminus\{b\}$.
  $\Lambda$, $\Lambda_a$ and $\Lambda_b$ denote the induced
  adjacency relations. In addition we require
  \begin{enumerate}
  \item system $\mathcal{L}$ to be stricter than the other systems, 
    i.e. $\Lambda \succcurlyeq \Lambda_a$ and  $\Lambda \succcurlyeq \Lambda_b$.
  \item that there is a leaf-graph $G_{ab}$, 
    describing an LC-equivalent graph state of system $\mathcal{L}$,
    such that $G_{ab}-a$ and $G_{ba}-b$ describe LC-equivalent graph states of $\mathcal{L}_a$ and
    $\mathcal{L}_b$, respectively.
  \item the surface code systems $\mathcal{L}_a$ and $\mathcal{L}_b$ to be nonlocal.
  \end{enumerate}
  Then the surface code system $\mathcal{L}$ is nonlocal.
\end{Theorem}

\begin{figure}[tbp]
  \centering
  \includegraphics[width=1.0\linewidth]{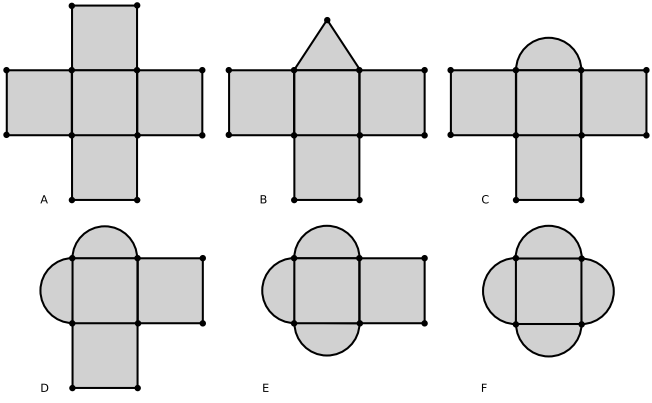}  
  \caption{\label{fig:example_sys5}
    Example for the reduction procedure based upon Theorem~\ref{theo:reduction}.
    \textbf{A:} The $+$-shaped pentomino as presented in Fig.~\ref{fig:mintcm}~(A).
    \textbf{B-C:} By reducing the upper square twice one obtains successively smaller systems.
    \textbf{D-F:} This procedure is repeatedly applied to the remaining squares and leads to the
    system depicted in Fig.~\ref{fig:mintcm}~(C). The ``removal'' of each edge/qubit is justified by
    Theorem~\ref{theo:reduction}; thus the non-locality of the smallest setup (\textbf{F}) implies
    the non-locality of the whole chain of larger setups.
  }
\end{figure}

\begin{proof}
  In order to show that the surface code system defined by $\mathcal{L}$ is non-local, one 
  has to find at least one non-local edge for each element in $\left[G_{ab}\right]_\mathrm{LC}$
  with respect to the adjacency relation $\Lambda$. Let $H\in \left[G_{ab}\right]_\mathrm{LC}$
  be an arbitrary graph representing an LC-equivalent graph state. According to 
  Proposition~\ref{prop:sub}, each element in $\left[G_{ab}\right]_\mathrm{LC}$ has 
  at least one subgraph in $\left[G_{ab}-a\right]_\mathrm{LC}$
  or $\left[G_{ba}-b\right]_\mathrm{LC}$. W.l.o.g. 
  let $H^\prime_a\in\left[G_{ab}-a\right]_\mathrm{LC}$ be such
  a subgraph, i.e. $H^\prime_a\subseteq H$. Since $H^\prime_a$ describes an
  LC-equivalent graph state of system $\mathcal{L}_a$ (see 2.) there is at least one edge 
  $e=\{v_1,v_2\}\in \mathbb{E}\left(H^\prime_a\right)\subseteq \mathbb{E}(H)$, 
  such that the vertices are not adjacent with respect to $\Lambda_a$, i.e. 
  $e\notin \mathbb{E}\left(G_{\Lambda_a}\right)$ (see 3.).
  Since system $\mathcal{L}$ is stricter than $\mathcal{L}_a$ (see 1.) it follows 
  $G_{\Lambda}\left[\mathbb{E}(\mathcal{L}_a)\right]\subseteq G_{\Lambda_a}$ and consequently 
  $e\notin \mathbb{E}\left(G_{\Lambda_a}\right) \Rightarrow e\notin \mathbb{E}\left(G_{\Lambda}\left[\mathbb{E}(\mathcal{L}_a)\right]\right)$.
  Since $v_1,v_2\in \mathbb{E}(\mathcal{L}_a)$ it follows furthermore 
  $e\notin \mathbb{E}\left(G_\Lambda\right)$.
  Thus we found an edge $e$ in $\mathbb{E}(H)$ that represents a non-local 
  interaction in the larger system $\mathcal{L}$. 
  Since $H$ was chosen arbitrarily, this shows the non-locality of $\mathcal{L}$.
\end{proof}

Using Theorem \ref{theo:reduction}, the $+$-shaped pentomino (see Fig.~\ref{fig:mintcm}~(A)) 
can be reduced to the $8$-qubit setup depicted in Fig.~\ref{fig:mintcm}~(C). 
The procedure applied in this particular case is depicted in Fig.~\ref{fig:example_sys5}. 
The peripheral edges are contracted successively until we end up with the $8$-qubit 
setup in Fig.~\ref{fig:example_sys5}~(F). The contractions are justified by
Theorem~\ref{theo:reduction}. Each step requires an LC-equivalent graph state with 
leaf-graph $G_{ab}$ as depicted in Fig.~\ref{fig:mintcm}~(A), such that $G_{ab}-a$ 
describes an LC-equivalent graph state of the next (reduced) setup. One can perform 
these steps easily by hand (they are not very exciting!) and, as a consequence, verify
the non-locality of the $+$-shaped pentomino using the non-locality of the $8$-qubit setup.

\end{document}